\documentclass[a4paper,11pt]{article}
\pdfoutput=1
\usepackage{jcappub}
\usepackage[latin1]{inputenc}
\usepackage{amsmath}
\usepackage{amsfonts}
\usepackage{amssymb}
\usepackage{graphicx}
\usepackage{dcolumn}
\usepackage{caption}
\usepackage{float}
\usepackage{bm}
\usepackage{latexsym}
\usepackage[sort&compress]{natbib}
\usepackage[normalem]{ulem}
\usepackage{graphicx}
\usepackage{subcaption}
\usepackage[colorlinks=true]{hyperref}
\usepackage{multirow}
\usepackage{array}

\newcommand{\Ne}{n_{\rm e}}
\newcommand{\NH}{n_{\rm H}}
\newcommand{\NHe}{n_{\rm He}}
\newcommand{\oih}{{}^{\rm H}\omega_{i}}
\newcommand{\op}{\omega_{\rm p}}
\newcommand{\oihe}{{}^{\rm He}\omega_{i}}
\newcommand{\fih}{f_i^{\rm H}}
\newcommand{\fihe}{f_i^{\rm He}}

\newcommand{\me}{m_{\rm e}}
\newcommand{\meff}{m_{\rm eff}}
\newcommand{\mg}{m_{\rm \gamma}}
\newcommand{\ma}{m_{\rm a}}
\newcommand{\Dosc}{\Delta_{\rm osc}}
\newcommand{\losc}{\ell_{\rm osc}}

\newcommand{\De}{\Delta_{\rm e}}

\newcommand{\Da}{\Delta_{\rm a}}
\newcommand{\Daga}{\Delta_{\rm \gamma a}}
\newcommand{\Pga}{P({\rm \gamma \rightarrow a})}
\newcommand{\gammaa}{{\rm \gamma a}}
\newcommand{\gammaad}{{\rm \gamma_{ad}}}
\newcommand{\BT}{B_{\rm T}}

\newcommand{\Rv}{R_{\rm V}}
\newcommand{\kB}{k_{\rm B}}

\title{Polarized anisotropic spectral distortions of the CMB: Galactic and
  extragalactic constraints on photon-axion conversion}
\author[a,b,c] {Suvodip Mukherjee,}\emailAdd{smukherjee@flatironinstitute.org}
\author[d] {Rishi Khatri}\emailAdd{khatri@theory.tifr.res.in}
\author[a,b,c,e]{and Benjamin D. Wandelt}\emailAdd{wandelt@iap.fr}
\affiliation[a]{Center for Computational Astrophysics, Flatiron Institute, 162 5th Avenue, 10010, New York, NY, USA}
\affiliation[b]{Institut d'Astrophysique de Paris\\ 98bis Boulevard Arago, 75014 Paris, France}
\affiliation[c]{Sorbonne Universités, Institut Lagrange de Paris \\ 98 bis Boulevard Arago, 75014 Paris, France}
\affiliation[d]{Tata Institute of Fundamental Research\\ Homi Bhabha Road, Mumbai, 400005, India}
\affiliation[e]{Departments of Physics and Astronomy, University of Illinois at Urbana-Champaign, 1002 W Green St, Urbana, IL 61801, USA}
\date{\today}
\keywords{Axions, Cosmic Microwave Background, Spectral Distortions}

\abstract{\noindent {We revisit the cosmological constraints on resonant
    and non-resonant conversion of photons to axions in the cosmological
    magnetic fields. We find that the constraints on photon-axion coupling
    and primordial magnetic fields are much weaker than previously claimed
    for low mass axion like particles with masses $\ma \lesssim 5\times
    10^{-13}\,\text{eV}$. In particular we find that the axion mass range $10^{-14}\, \text{eV} \le \ma \le 5\times
10^{-13}\, \text{eV}$ is not excluded by  the CMB data contrary to the previous claims. We also examine the photon-axion conversion in the Galactic
    magnetic fields. Resonant conversion in the large scale coherent Galactic magnetic field
    results in  $100\%$ polarized anisotropic spectral distortions of the  CMB for the mass range $10^{-13}\, \text{eV} \lesssim \ma
  \lesssim 10^{-11}\, \text{eV}$. The polarization pattern traces the transverse to
  line of sight component of the Galactic magnetic field while both the
  anisotropy in the Galactic magnetic field and electron distribution
  imprint a characteristic anisotropy pattern in the spectral
  distortion. Our results apply to scalar as well as pseudoscalar
  particles. For conversion to scalar particles, the polarization
    is rotated by $90^{\circ}$ allowing us to distinguish them from the pseudoscalars. For $\ma \lesssim 10^{-14}\,\text{eV}$ we have non-resonant conversion
  in the small scale turbulent magnetic field of the Galaxy
  resulting in anisotropic but unpolarized spectral distortion in the CMB.  These unique signatures are potential
  discriminants against the isotropic and non-polarized signals such as
  primary CMB, and $\mu$ and $y$ distortions with the anisotropic nature
  making it accessible to experiments with only relative calibration like
  Planck, LiteBIRD, and CORE. We forecast for PIXIE as well as
  for these experiments using Fisher matrix formalism.}} 
\begin{document}
\maketitle

\pagenumbering{arabic}
\thispagestyle{plain}
\markright{}

\section{Introduction}
The Cosmic Microwave
  Background (CMB) was discovered by Penzias \& Wilson \cite{pw1965} and later found to have
 { an almost perfect blackbody spectrum by} Far Infrared Absolute
 Spectrophotometer (FIRAS) experiment \cite{firas,firasan,fm2002,2009ApJ...707..916F} with
 temperature $2.7255$ K (almost simultaneously confirmed by the rocket based
 experiment COBRA \cite{ghw1990} with slightly less sensitivity.) {Another instrument on board COBE, the differential microwave
 radiometer (DMR), discovered the nearly statistically isotropic
 fluctuations of order $10^{-5}-10^{-4}$ K on top of the $2.7$K
 background. The exquisite measurements of the CMB temperature and
 polarization angular} fluctuations over the
past few decades by several space-based (COBE \cite{cobe}, WMAP
\cite{Bennett:2012zja} and Planck \cite{Adam:2015rua}), ground-based (SPT
\cite{Hanson:2013hsb}, ACT \cite{actpol}, BICEP-KECK \cite{Array:2016afx},
POLARBEAR \cite{Ade:2013gez} etc.) and balloon-based (BOOMERANG
\cite{boomerang}, MAXIMA \cite{Stompor:2003kr}, etc.) missions are well
explained by the $6$ parameter $\Lambda$CDM (Lambda Cold Dark Matter) model
and is one of the fundamental pillars of the standard cosmological
model. Along with the {angular fluctuations of the CMB} field, {we also
expect deviations from the blackbody spectrum within the standard
cosmological scenario} \cite{zeldovich,sz1970, Chluba:2011hw, Chluba:2012gq,
  Khatri:2012tv, Khatri:2012rt, Khatri:2012tw, sk2013,Chluba:2016bvg,
  Hill:2015tqa, Emami:2015xqa} and measurement of these would deepen our
understanding of  {both early} and late time epoch of the Universe. {Only one
type of spectral distortion, the Sunyaev-Zeldovich effect or the $y$-type
distortion \cite{zeldovich}, has so far been detected towards the clusters of galaxies \cite{Hasselfield:2013wf,Bleem:2014iim, Ade:2015gva, Staniszewski:2008ma, 2010A&A...518L..16Z, 2013A&A...550A.134P}.
} 
   Spectral distortions in CMB {are} also predicted by several high energy physics
scenarios which are important particularly in the pre-recombination
epoch \cite{Tashiro:2012nb,Tashiro:2012pp, Chluba:2009uv, Colafrancesco:2004sp, Tashiro:2008sf, Pani:2013hpa,Blum:2016cjs}.  In brief, {CMB
  spectral distortions provide an unexplored and extremely rich window to several astrophysical
and cosmological phenomenon.} 

{The current best constraints on the deviation of the sky-averaged
  CMB (monopole) from a blackbody spectrum come from FIRAS
\cite{firas,fm2002}  which gave an}
upper bound on spectral distortions of {$\Delta I_\nu/I_\nu \lesssim
5\times 10^{-5}$ at the peak of the blackbody spectrum}. Recently, the constraints on the anisotropic spectral distortions,
including the fluctuating contribution to the  {all sky average} $y$ -distortion were
obtained from the Planck and SPT data in \cite{Khatri:2015jxa, Khatri:2015tla}. Upcoming proposed CMB missions
like PIXIE \cite{Kogut:2011xw} {would have} an instrumental noise of nearly $3-4$
orders of magnitude better than FIRAS and hence {would be able to}
measure the CMB spectral
distortions with an unprecedented accuracy \cite{2013JCAP...06..026K}.  {Measurement of the spectral distortions signal will also depend upon the successful cleaning of the foreground contaminations \cite{Khatri:2015tla, Khatri:2015gxa, Aghanim:2015eva, Remazeilles:2018kqd, Abitbol:2017vwa}. } Other CMB missions like CORE\cite{2017arXiv170604516D} and LiteBIRD
\cite{Matsumura:2016sri} would be able to measure the spatially fluctuating part
of the spectral distortions \cite{Tashiro:2014pga} at a much {better
precision compared to} Planck. These experiments would be
polarization sensitive in all frequency channels. These missions {could} pioneer a new era in cosmology by measuring
several guaranteed but unexplored cosmological imprints {on the CMB spectrum.}
 Along with
several well known sources of spectral distortions (such as $\mu$ \cite{sz1970,
  Chluba:2011hw, Khatri:2012tw}, $y$ \cite{Hill:2015tqa, Khatri:2012tw,
  Chluba:2012gq}, dark matter annihilation \cite{Chluba:2011hw}), coupling
between photons and  {pseudoscalar axion like particles (ALPs) or light
scalar particles (LSPs) in the} presence of external magnetic field
\cite{axion_1, sikivie, raffelt_1, anselm, raffelt,raffeltbook, axion_2,
  Tashiro:2013yea} is also a potential source of spectral distortion. 
ALPs are one of the promising candidates for dark
matter and may be a solution to some of the anomalies in the standard
$\Lambda$CDM model \cite{turner1983,prs1990,sin1994,hbg2000,scb2014,hotw2017}. Regardless of whether the ALPs  form the bulk of the dark
  matter, they are predicted almost ubiquitously in many beyond standard
  model theories of particle physics, including the string theory \cite{db2009,ringwald2014,marsh2016}.  Indirect
astrophysical searches of ALPs  along with the growing ground-based
experimental efforts  \cite[see][for a review of ground based experiments]{axionreview} like CAST \cite{cast}, ALPS-II \cite{alpsii}, MADMAX
\cite{madmax},  ADMX \cite{admx},  CASPER \cite{casper} are therefore very important.

The magnetic field is present in the Universe at different scales with a
varying strength. The extragalactic magnetic field is expected to
  be  of the order of $10^{-9}$ Gauss (nG) or smaller
  \cite{avw2014}. The extragalactic magnetic fields, particularly
    in voids and in the early Universe, if primordial in origin, 
  must be stochastic (Gaussian random fields), described by a power spectrum
  (possibly scale invariant) \cite{primordialB}, and therefore  without a single coherence
  scale. Previous studies \cite{axion_2,Tashiro:2013yea} have considered
Mpc scale extragalactic magnetic fields to study the imprints of the
photon-ALP  {or photon-LSP} conversion on CMB photons. 
The magnetic field is however known to be  present at the Galactic (kiloparsec) scales (kpc
scales), compared to only upper limits on the primordial intergalactic
magnetic fields \cite{Ade:2015cva}, with much better understanding of its strength
  and  morphology \cite{Jansson:2012rt,jansson}. 

In this paper, we focus on an unexplored scenario concerning {the}
spectral distortions of CMB photons due to photon-ALP  {or photon LSP} conversion in
the presence of the local magnetic field from Milky Way.  {We will
consider the  ALPs from now on for definiteness but our results are
applicable to LSPs also in a straightforward way as explained in Sec. \ref{sec:lsp}.} The CMB photons
passing through the Galactic halo to reach the earth can get converted to
ALPs in the presence of the $10^{-6}$ Gauss ($\mu\, G$) magnetic field
depending upon photon-ALP coupling $g_{\gammaa}$, ALP mass $\ma$,
electron ($\Ne$) and neutral hydrogen ($\NH$) number densities and strength of the magnetic field. As the
Galactic magnetic field is not isotropic {but exhibits large scale
  coherent structure and small-scale turbulent fluctuations} \cite{jansson,
  Jansson:2012rt, Oppermann:2011td, 2016A&A...596A.103P}, the spectral
distortions induced by the photon-ALP conversion must also exhibit
{large scale anisotropy  and should be correlated with the large
  scale structure in the Galactic magnetic field.} In particular,
the regions of the sky with the stronger magnetic field can convert photons
to ALPs more efficiently and than the parts of the sky with the weaker
magnetic field. Secondly, the presence of fluctuations in the spectral
distortion makes this phenomenon measurable from CMB missions like Planck
and LiteBIRD which have only relative calibration. 
{The spectral distortion signal from the photon-ALP conversion exhibits a unique structure in both frequency and spatial domain, which makes it easier to distinguish from other cosmological (or astrophysical) sources and other systematics.}

We review the
physics of   photon-axion conversion and re-evaluate
  the existing cosmological constraints on photon-ALP conversion in section
\ref{review}. We discuss the
signatures of spectral distortion due to photon-ALP conversion in the Milky
Way in
Secs. \ref{spectral-dist-reso} and \ref{non-resonant} and forecast the measurability of this
phenomenon from several CMB missions like Planck, PIXIE, LiteBIRD, and CoRE
in Sec. \ref{forecast} using Fisher Matrix.  Finally in
Sec. \ref{conclusion}, we conclude our study and discuss its future
implications. We use natural units (with reduced Planck constant, speed of light
and Boltzmann constant respectively set to unity $\hbar=c=\kB=1$) when discussing
physics but restore physical constants when discussing observations.

 
\section{Review of CMB-ALP conversion physics and current constraints
  from cosmology}\label{review}
Photon-ALP conversion and its cosmological consequences are
  well studied topics in the literature \cite{sikivie, raffelt_1, anselm,
  raffelt, Vysotsky:1978dc, Berezhiani:1992rk, Jain:2002vx, Millea:2015qra, Vogel:2017fmc, PhysRevD.76.121301, PhysRevD.84.105030, PhysRevD.86.085036, Schlederer:2015jwa, DAmico:2015snf}.  
ALPs  and photons oscillate into each other in the presence
of a magnetic field \cite{sikivie, raffelt_1, anselm, raffelt,axion_2}. {The
interaction is given by 
\begin{align}
\mathcal{L}_{\rm int}=g_{\gammaa}\mathbf{E_{\gamma}.B_{\rm ext}}a,\label{Eq:int}
\end{align}
where $\mathbf{B}_{\rm ext}$ is the external magnetic field, $\mathbf{E}_{\gamma}$ is the
electric field of the photon, $a$ is the axion field and $g_{\gammaa}$ is the photon-ALP coupling. Thus only the polarization with its electric
field aligned with external magnetic field couples to the axion. Obviously
photons and axions will couple only if the magnetic field has a component
$\mathbf{B_T}=\mathbf{B}_{\rm ext}-\left(\mathbf{B}_{\rm ext}.\mathbf{\hat{k}}\right)\mathbf{\hat{k}}$
transverse to the photon propagation direction $\mathbf{\hat{k}}$.
}
The evolution equation for the two state quantum system,
assuming relativistic ALP, is
given by \cite{raffelt_1,raffeltbook}
\begin{align}
\left(\omega+\left(
\begin{array}{cc}
\De & \Daga\\
\Daga & \Da\\
\end{array}
\right)+i\partial_z\right)
\left(
\begin{array}{c}
A_{\parallel}\\
a\\
\end{array}
\right)=0\label{Eq:evol}
\end{align}
Here we want to study the evolution of the system along a spatial direction
$z$, $\omega$ is the temporal frequency, in Fourier space 
$i\partial_z\rightarrow k$ the spatial frequency or the momentum and $A_{\parallel}$ is the
photon polarization that is parallel to the component of the magnetic field
$\mathbf{B_T}$ transverse to the propagation direction. The mixing
matrix elements are defined below. {In general $\De$, which is a
  function of free electron, atomic and molecular densities, will vary
  along the photon geodesic as CMB photons travel cosmological
  distances. $\Daga$ is a function of magnetic field which is also
  spatially varying. We will, at first, ignore these complications and look at the
  solutions in the presence of homogeneous medium and magnetic fields and
  return to  {the} inhomogeneous case later.}  The equation \ref{Eq:evol}  is solved by diagonalizing
the $2\times 2$ matrix on the left hand side through rotation by mixing
angle $\theta$.
 The probability of conversion of a photon (with linear
polarization parallel to the component of magnetic field transverse to the
propagation direction) to an ALP is then given {in the homogeneous case by} \cite{axion_2}
\begin{align}\label{prob}
P(\gamma \rightarrow a) & = \frac{(\Delta_{\gammaa}s)^2}{(\Dosc s/2)^2}
\sin^2(\Dosc s/2) \nonumber\\
 &\equiv  \sin^2(2\theta)\sin^2(\Dosc s/2) 
\end{align}
where $\BT$ is the transverse {(to photon momentum)} component of
the magnetic field, $s$ is the distance
 travelled by the photons, $\theta$ is the mixing angle defined so that 
  $2\theta \rightarrow 0$ at high electron densities,
\begin{align}
\cos(2\theta) &=\frac{\Da-\De}{\Dosc}\\
\Dosc^2 =& (\Da  - \De )^2 +4\Delta^2_{\gammaa},
\end{align}
,
\begin{align}
\De\equiv (n-1)\omega,
\end{align}
and $n$ is the refractive index for photon propagation through matter. {As
we see below, for high electron densities photon has real effective mass,
$\De \propto -\mg^2 < 0$ and $\Da \propto
-\ma^2 < 0$ and in the limit
$|\De| \gg |\Da|,|\Daga|$ mixing is suppressed ($\sin2\theta\rightarrow 0$)
and we get
$\cos(2\theta)\rightarrow 1$. Similarly in vaccum, when $\De\rightarrow 0,
|\Da|\gg |\Daga|$, again the mixing is suppressed ($\sin2\theta\rightarrow 0$) but we have
$\cos2\theta\rightarrow -1$. Photon axion mixing is maximum at resonance, $\De=\Da$
giving $2\theta=\pi/2, \sin2\theta=1$.} 

For
astrophysical matter densities, $|n-1|\ll 1$ and can be approximated (away
from resonance frequencies $\omega_i$ of atoms) by \cite{born1999}
\begin{align}
n-1 \approx \frac{-\mg^2}{2\omega^2}\approx \frac{2\pi\alpha}{\me\omega^2}
\left(-\Ne +\NH \sum_i
\frac{\fih \omega^2}{\oih^2-\omega^2}+\NHe\sum_i\frac{\fihe \omega^2}{\oihe^2-\omega^2}\right),
\end{align}

where $\mg$ is the effective mass of the photon, $\alpha$ is the fine
structure constant, $\omega$ is the angular frequency, $\Ne,\NH$ and $\NHe$ are the number
the density of free electrons, neutral hydrogen, and neutral helium respectively,
$f_i^{\alpha}$ is the oscillator strength of element $\alpha\in\{{\rm
  H,He}\}$  for transitions with energy ${}^{\alpha}\omega_i$ from the
ground state. For free electrons, there is no resonant frequency and
 the effective mass squared is positive. For neutral atoms, such
as hydrogen, the effective mass squared is negative below the resonant
frequency. Neutral atoms exist only after recombination and in ground state
with the minimum resonant frequency of $10.2$ eV corresponding to
Ly-$\alpha$ transition of hydrogen. For CMB we will therefore always have $\omega
\ll \omega_i$ and we can approximate $\De$ as (ignoring
helium and heavier elements) \cite{mirizzi2009}
\begin{align}
\De&\approx   \frac{\op^2}{2\omega}
\left[-1 +7.3\times 10^{-3}\frac{\NH}{\Ne}  \left(\frac{\omega}{{\rm
        eV}}\right)^2\right]\nonumber\\
&=-2.6\times
10^6 \left(\frac{\Ne}{10^{-5}{\rm cm^{-3}}}\right)\left(\frac{100~{\rm
    GHz}}{\nu}\right)\left[1 -7.3\times 10^{-3}\frac{\NH}{\Ne}  \left(\frac{\omega}{{\rm
        eV}}\right)^2\right]{\rm Mpc}^{-1}\label{Eq:de}
\end{align}
where $\op^2=4\pi \alpha\Ne/(\me)$ is the plasma frequency.

\begin{figure}[h]
    \centering
    \begin{subfigure}[b]{0.47\textwidth}
        \includegraphics[width=\textwidth]{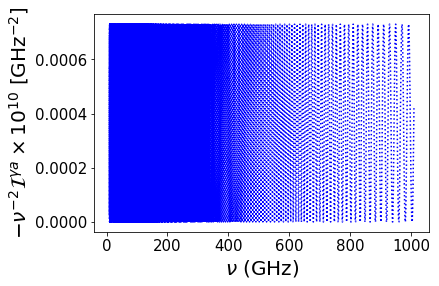}
        \caption{$s= 1$ kpc}
        \label{fig:nu}
    \end{subfigure}
    ~ 
    \begin{subfigure}[b]{0.45\textwidth}
        \includegraphics[width=\textwidth]{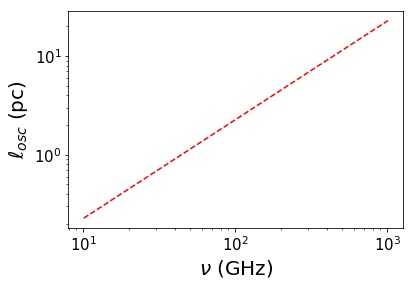}
        \caption{$n_e= 10^{-5}$ cm$^{-3}$}
        \label{fig:s}
    \end{subfigure}
       \caption{{The left panel shows frequency dependence of the CMB
           spectral distortion  with $\Ne=
           10^{-5}\text{cm}^{-3}$ and $g_{10}B_{\mu\text{G}}=1$.  The right
           panel shows oscillation length $\losc$ as a function of photon
           frequency for same parameters. The axion mass is assumed to be
           small compared to the effective photon mass. }\label{spectra}}
\end{figure}

{For the range of parameters of interest we also have
\begin{align}
\Delta_{\gammaa} &\equiv \frac{g_{\gammaa} |\BT|}{2} &=& ~15.2
\left(\frac{g_{\gammaa}}{10^{-11}{\rm Gev}^{-1}}\right)\left(\frac{\BT}{\mu{\rm G}}\right){\rm{Mpc^{-1}}},&\label{Eq:dg}\\
\Da &\equiv - \frac{m^2_a}{2\omega}&=&-1.9\times
10^4\left(\frac{\ma}{10^{-14}{\rm eV}}\right)\left(\frac{100~ {\rm
      GHz}}{\nu}\right){\rm Mpc}^{-1}&\label{Eq:da}
\end{align}
}
$g_{\gammaa}$ is the photon-ALP coupling and $\omega=2\pi \nu$. The photon polarization state orthogonal to
 $\BT$ is unaffected. Thus initially unpolarized light propagating
 through a magnetic field will become polarized as intensity in one of the linear
 polarizations is decreased due to photon-ALP oscillation. These results
 apply only if both the electron density and the magnetic field are
 homogeneous. The departure from BlackBody (BB) spectrum for the
   affected polarization can be quantified by 
\begin{equation}\label{spec1}
\mathcal{I}^{\gammaa}=\frac{\Delta I^{\gammaa}_\nu}{I_\nu} \equiv \frac{I^{\text{obs}}_\nu -I_\nu}{I_\nu}= -\bar{P}(\gamma \rightarrow a),
\end{equation}
where, $I_\nu= \left(h\nu^3/c^2\right)/(e^{h\nu/\kB T_{\text CMB}}-1)$ is the
Planck spectral form for single polarization in physical units, $h$ is the Planck's constant,
$c$ is the speed of light and $\kB$ is Boltzmann constant. This is
plotted in Fig. \ref{fig:nu}. We see that it is a fast oscillating function
of frequency (as well as  distance $s$) due to the second factor in Eq. \ref{prob}. In any
experiment with reasonable frequency and angular resolution we will only
detect the result of average over many oscillations. We can therefore
replace the oscillating factor with its average over an oscillation giving 
\begin{equation}\label{prob2}
P(\gamma \rightarrow a) = \frac{2\Delta^2_{\gammaa}}{\De^2},
\end{equation}
if the magnetic field is nearly coherent over the
scales of size $s\gg \losc \equiv 2\pi/\Dosc$. The oscillation length,
$\losc$ is plotted in Fig. \ref{fig:s} as a function of frequency $\nu$.

In reality both the electron density and magnetic fields are
inhomogeneous. In particular the primordial magnetic fields (because of
stochastic initial conditions) as well as
small scale Galactic magnetic fields (because of turbulence)  are expected
to be stochastic. In this case, as a toy model,  we can approximate the magnetic fields as
composed of independent domains of size $d_0$ such that the magnetic field {and electron density are}
 homogeneous inside each domain but in different domains the magnetic
field has different random
orientations but same strength for simplicity. In the limit of large number
of domains we can obtain an analytical solution for the conversion
probability of the total unpolarized intensity given by \cite{grossman, axion_2}
\begin{equation}\label{prob-1}
\bar{P}(\gamma \rightarrow a) (r) = \frac{1}{3}\bigg(1-e^{(-3P(\gamma
  \rightarrow a)r/2d_0)}\bigg)\, \hspace{2cm} r>>d_0,
\end{equation}
where $r$ is the size of the turbulent region, $\Pga$ is the probability of conversion inside each domain and
  $\bar{P}$ is the average conversion probability over the whole turbulent
  region.
In this case,  we  have
a spectral distortion but no polarization. 
In the limit $r \rightarrow \infty$ we saturate the probability with $1/3^{\text{rd}}$ of
the photons getting converted into ALPs.

The true situation will be in between the above two extreme limits and both the   magnetic field and electron density would vary
  along the photon geodesic.  In particular, when photons 
  and ALPs propagate through the inhomogeneous medium, there is the possibility
  of resonance when the effective mass of photon becomes equal to the mass
  of the ALP,
\begin{align}
\De = \Da.
\end{align} 
For  inhomogeneous matter distribution and magnetic fields, relevant for
  considering the CMB - ALP conversions, the conversion probability is
  sensitive to how fast the matter/electron density and the magnetic fields,
  and therefore the mixing angle $2\theta$, 
  change compared to the oscillation length, $\losc=2 \pi /\Dosc$
  \cite{raffeltbook}. We therefore define an adiabaticity parameter (with
  $\nabla$ denoting the spatial derivative with respect to the physical
  distance along the line of sight or proper time),
\begin{align}
\gammaad &= \left|\frac{\pi}{\losc \nabla \theta}\right|\nonumber\\
&=\left|\frac{\Dosc}{\sin(2\theta)\cos(2\theta)\nabla(\ln\Daga)+\sin(2\theta)\De/\Dosc\nabla(\ln\De)}\right|\label{Eq:gad}
\end{align}
with the adiabatic limit defined as $\gammaad \gg 1$. The propagation is
adiabatic when the length scale over which the mixing angle changes,
$1/\nabla \theta$ is much larger than the oscillation length $\losc$. In
the adiabatic limit, the final conversion probability depends only on the
initial mixing angle when the photon is emitted, $\theta_0$, and the final
mixing angle at the detector $\theta$ irrespective of whether there is a resonance  \cite{raffeltbook}{
\begin{align}
\Pga=\frac{1}{2}\left(1-\cos 2 \theta_0\cos 2 \theta\right)\label{Eq:ad}
\end{align}
}

\begin{figure}[H]
\centering
\includegraphics[width=4.2in,keepaspectratio=true]{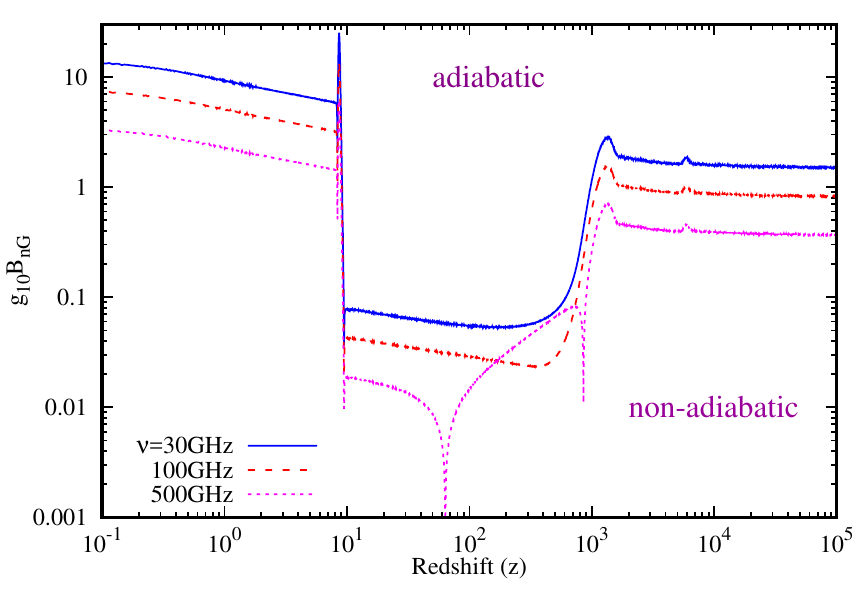}
\captionsetup{singlelinecheck=on,justification=raggedright}
\caption{Transition condition between the adiabatic and non-adiabatic
  resonance is plotted as $g_{\gammaa}B_{\rm T}$(comoving) in
  units of $10^{-10}~{\rm GeV^{-1}nG}$. The $g_{\gammaa}B_{\rm T}$ much
  larger than the threshold curves will result in adiabatic resonances
  while much smaller values will result in non-adiabatic resonances.}\label{Fig:adiab}
\end{figure}

The first term in the
denominator in Eq. \ref{Eq:gad} arises due to the inhomogenous magnetic fields and the second
term due to inhomogeneity in the matter distribution and ionization
fraction. At resonance $\sin(2\theta)=1;\cos(2\theta)=0$ and the first term
in the denominator due to the inhomogeneous magnetic field
vanishes. Therefore, for resonant conversion only the inhomogeneity in the
matter is relevant and the adiabaticity parameter becomes
\begin{align}
\gammaad(\rm resonance) &= \frac{4\Daga^2}{|\nabla \De|}.
\end{align}
A given cosmological average recombination and reionization history fixes the
denominator. We can therefore plot the $g_{\gammaa}B_{\rm T}$ for which
$\gammaad(\rm resonance)=1$ as a function of redshift. This curve separates
the adiabatic and non-adiabatic regions of the parameter space at each redshift. For values of
$g_{\gammaa}B_{\rm T}$ above this  curve we will have adiabatic resonances
for ALP mass which satisfies $\ma=\mg$ at that redshift and below this
curve we will have a non-adiabatic resonance. This is shown in
Fig. \ref{Fig:adiab}. The comoving magnetic field is plotted which is
related to the physical magnetic field by $\BT(\rm physical)=(1+z)^2\BT(\rm
comoving)$.

\begin{figure}
\centering
\includegraphics[width=4.2in,keepaspectratio=true]{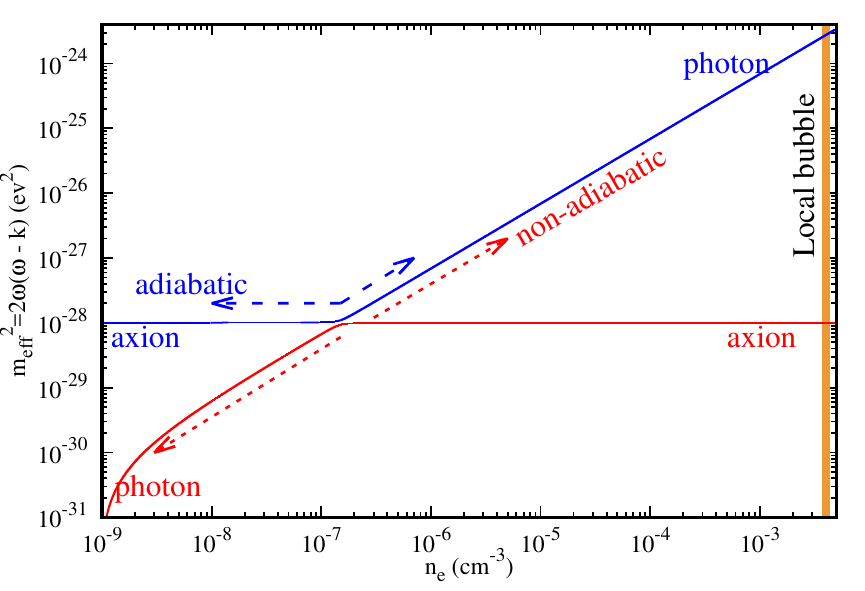}
\captionsetup{singlelinecheck=on,justification=raggedright}
\caption{Dispersion relations for the photon-ALP system as a function of
  the electron density. The electron density in the solar neighbourhood, in
particular the local bubble, is also marked. For this plot the ALP mass is
$\ma=10^{-14}~{\rm eV}$, and $g_{\gammaa} \BT=10^{-6}{\rm Gev^{-1}nG}$. Also
shown are the trajectories along the dispersion relations for adiabatic and
non-adiabatic cases when photons/ALP are propagating thought an
inhomogeneous plasma.}\label{Fig:disp}
\end{figure}

\begin{figure}
\centering
\includegraphics[width=4.2in,keepaspectratio=true]{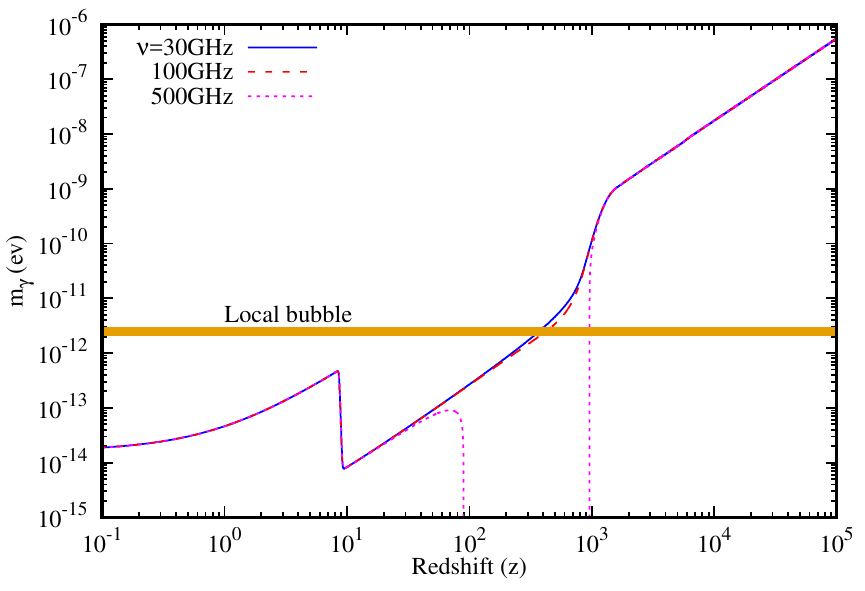}
\captionsetup{singlelinecheck=on,justification=raggedright}
\caption{Cosmic evolution of the photon effective mass $\mg$. We assume a
  standard $\Lambda$CDM recombination history  with reionization happening
  at $z=8$ calculated using CLASS \cite{class} in HyREC mode \cite{hyrec2,hyrec,cosmorec}. }\label{Fig:mgamma}
\end{figure}

For the non-trivial solutions of Eq. \ref{Eq:evol} to exist, the
determinant of the operator on the left hand side must vanish. This gives
us two dispersion relations \cite{raffeltbook} 
 defining the two eigenstates $(\meff^2=\omega^2-k^2\approx 2\omega
 (\omega-k))$ of the system,
\begin{align}
2\omega(\omega-k)&=-\omega\left(\De+\Da\right)\pm \omega\Dosc\nonumber\\
&=\frac{\ma^2+\mg^2}{2}\pm\left[\left(\frac{\ma^2-\mg^2}{2}\right)^2+\omega^2g_{\gammaa}^2B_{\rm T}^2\right]^{1/2}
\end{align}
The dispersion relations or the eigenstates of the Hamiltonian are shown in Fig \ref{Fig:disp} as a function of
the electron number density assuming fully ionized plasma. 
The resonance
happens when $\ma=\mg$. If the resonance is adiabatic, $\gammaad\gg 1$, the
system stays in the instantaneous eigenstate on the same branch of the 
dispersion relation. In particular in this case a
photon produced at high density away from the resonance
 would follow the
upper branch as the density of the medium decreases and we will have a full conversion to ALPs at sufficiently
low densities\footnote{{This is similar to the MSW effect \cite{ms1985,w1978,b1986} in case of neutrino
  flavor oscillations}}. This can also be seen from  Eq. \ref{Eq:ad} with
{$\cos 2 \theta_0\approx 1,\cos 2\theta\approx -1$}  giving $\Pga \approx 1$. If the density of the medium changes rapidly compared to the
oscillation length, $\losc$, there is a non-zero probability that the
quantum system would make a transition between the two eigenstates
 or branches of the dispersion relation. In the limit that the change in
 density near the resonance is linear, the transition probability is given
 by the Landau-Zener formula
 \cite{landau1932,zener1932,stueckelberg1932,kp1989,raffeltbook} 
\begin{align}
p=e^{-\pi \gammaad/2}
\end{align}
and the conversion probability is
\begin{align}
P(\gamma \rightarrow a) =\frac{1}{2}\left(1-(1-2p)\cos 2\theta_0\cos 2\theta\right)\label{Eq:level}
\end{align}
The cosmic evolution of the photon effective mass $\mg$ is shown in
Fig. \ref{Fig:mgamma}.
For detection at earth, we should take into account the fact that solar
system is inside a local hot bubble with electron density $\sim 5\times
10^{-3}{\rm cm^{-3}}$ \cite{cr1987,frs2011,scc2014} with radius 
$\sim 100$ pc. The ionization fraction is high enough that we can ignore
the neutral hydrogen contribution to the photon mass, yielding $\mg^{\rm local}\approx
2.6\times 10^{-12}~{\rm eV}$. This is also shown in
Fig. \ref{Fig:mgamma}. For ALP masses far from resonance, we have the
oscillation length $\losc \lesssim  7\times 10^{-3}(\nu/150 {\rm GHz}){~\rm
  pc}$. Therefore, for $\ma \gg \mg^{\rm local}$ we are detecting the
photons in vaccum while for $\ma \ll \mg^{\rm local}$ we are detecting the
photons at high electron density. Previous analysis of cosmological CMB
constraints has assumed the cosmic average electron density of $\Ne\approx
2\times 10^{-7}~{\rm cm^{-3}}$ for calculating $\mg$ at the detection point
\cite{mrs2009,Tashiro:2013yea}. 

\subsection{Cosmological constraints: resonant case}
Before the epoch of electron-positron annihilation at $z\sim 10^8-10^9$,
the high electron-positron number density results in large scattering rate
of photons with electron and positrons,
heavily damping the photon-ALP conversions
\cite{Stodolsky1987,jrr2008}. We can therefore take as the initial state
pure photons at $z\sim 10^8$.
Depending on the ALP mass, we may have
one or more resonances between $z=10^8$ and today. We can roughly estimate
the effect of resonances on the final state from Fig. \ref{Fig:disp}. As
long as the resonances are non-adiabatic and the detection is done far from
the resonance, the probability of conversion remains small. Also, if there
are an even number of completely adiabatic resonances, we again end up with
a photon. To get a large conversion probability from photon to ALP we
must have an odd number of adiabatic resonances. From
Fig. \ref{Fig:mgamma}, we see that for $\ma\gtrsim 3\times 10^{-12}$ there
is only one resonance and the condition that this resonance should be
non-adiabatic gives an upper limit on $g_{\gammaa}\BT$
(Fig. \ref{Fig:adiab}). For $8\times 10^{-15} \lesssim \ma \lesssim 2\times
10^{-12}$ multiple resonances happen. In this case, the most stringent
constraints would come from the most adiabatic resonance which would occur
between recombination and reionization from Fig. \ref{Fig:adiab}. These
constraints have already been derived by \cite{mrs2009} and we will not
repeat them here.  It was claimed in \cite{Tashiro:2013yea} that the occurrence
of two non-adiabatic resonances places very stringent constraints and in
particular rules out the mass range between $10^{-14}\le \ma \le 5\times
10^{-13}$. However, we see from Fig. \ref{Fig:disp} that this cannot
happen. In fact starting at high density and going through two
non-adiabatic resonances, we will end up in the top dispersion relation in
Fig. \ref{Fig:disp} on the right/high-density side of the resonance. In
\cite{Tashiro:2013yea} however it was assumed that the photons are finally
detected in vacuum, implying an additional adiabatic resonance which is not
present in the formula (for two resonances) that they used making these constraints of \cite{Tashiro:2013yea} invalid. If
there was indeed an additional adiabatic resonance, it would be this
resonance which would provide the final constraint on $g_{\gammaa}\BT$ as were
derived in \cite{mrs2009}. 

We can make the above statements precise as follows. Let us denote
  the two eigenstates after $a^{\rm th}$ resonance (far from the resonance)
  plotted in Fig. \ref{Fig:disp} by normalized states $|\psi_1(a)\rangle, |\psi_2(a)\rangle$.  Each of
  these eigenstates $\psi_i$ is a superposition of photon and axion as
  determined by the mixing angle $\theta$. The level crossing probability $p$
  then denotes the probability of going from initial state $|\psi_i(0)\rangle$ to
  final state $|\psi_j(1)\rangle$ after crossing one resonance,
\begin{align}
\left|\langle\psi_i(1)|\psi_j(0)\rangle\right|^2=\left(
\begin{array}{cc}
1-p & p\\
p & 1-p\\
\end{array}
\right).
\end{align}
In case of $N$ resonances we get \cite{kp1989,flm2003}
\begin{align}
\left|\langle\psi_i(N)|\psi_j(0)\rangle\right|^2&=
\left|\langle\psi_i(N)|
  \cdots|\psi_k(a)\rangle\langle\psi_k(a)|\psi_l(a-1)\rangle\langle\psi_l(a-1)| \cdots \psi_j(0)\rangle\right|^2\nonumber\\
 = \prod_{a=1}^{N}\left(
\begin{array}{cc}
1-p_a & p_a\\
p_a & 1-p_a\\
\end{array}
\right)
&\equiv
\left(
\begin{array}{cc}
1-p & p\\
p & 1-p\\
\end{array}
\right),
\end{align}
 where $p$ is the final level crossing probability after $N$
  resonances which can be
  written as
\begin{align}
p=\frac{1}{2}\left(1-\prod_{a=1}^{N}\left(1-2p_a\right)\right).\label{Eq:levelp}
\end{align}
In the above calculation we have ignored the interference between different resonances and treated the level crossing probabilities as
  classical probabilities. This is justified if there is decoherence in the
wavefunction evolution \cite{k1993,dd2007}, for example, due to  propagation through the
stochastic primordial magnetic fields in-between the
resonances.  {Starting at high electron densities, 
in} case of $N$ even, we will be back on the {right} side of the resonance in
Fig. \ref{Fig:disp} while for $N$ odd we will end up on the left  {(or low
electron density)} side of
resonance. The photon to axion probability is equal to the probability
that starting with an approximately pure photon case on the upper eigenstate at
high electron densities 
we end up on the axion line far from the resonance. It is therefore given
by
\begin{align}
\Pga= \begin{array}{cc}
p &: N~{\rm even}\\
1-p &:N~{\rm odd}\\
\end{array}
\end{align}
We note that we do not have a choice to independently choose $N$ even or
odd and at the same time require final detection at high density or in
vacuum. Specifying one condition automatically fixes the other.

 For the frequency range $\nu\gtrsim 150~{\rm GHz}$, we will
  always have two resonances during the dark ages for small axion masses
  on account of the effective mass of photon getting dominated by neutral
  gas and becoming imaginary and then real again once reionization starts.  In this case of two resonances,  starting at upper left
  curve in Fig. \ref{Fig:disp}, the conversion to axion would occur if at
  one of the resonances we cross level (with probability $p_i$) but fail to do so at the other
  resonance (with probabilty $(1-p_j)$), where $p_i$ is the level crossing
  probability for $i$th resonance. We therefore have the total conversion
  probability for two resonances
\begin{align}
\Pga&=p=p_1(1-p_2)+p_2(1-p_1)\nonumber\\
\end{align} in agreement with Eq. \ref{Eq:levelp} and similar situation in \cite{Tashiro:2013yea}.   This
result is true for all axion masses, including $10^{-14}\le \ma \le 5\times
10^{-13}$. It is also clear that if one of resonances is more adiabatic
than the other (larger $(1-p_i)$), then that resonance will dominate the
overall probability. Using $\Pga=p$ (instead of $\Pga=1-p$ used by \cite{Tashiro:2013yea})
for this mass
range yields the correct constraint which is similar to the mass range just
outside this interval and much weaker than what is claimed in \cite{Tashiro:2013yea}.

\subsection{Cosmological constraints:  non-resonant case}
For axion mass $\ma \le 10^{-14}$ the only resonances are the ones during the dark ages when the effective mass of the
photons becomes imaginary for $\nu\gtrsim 150 ~{\rm GHz}$ (observed
frequency today). These constraints are considered in
\cite{Tashiro:2013yea}. In this case we can also expect competitive constraints from
non-resonant photon-axion oscillations in the stochastic primordial
magnetic fields in voids \cite{axion_2}. Previous
studies of non-resonant conversion have relied on the toy model of randomly oriented magnetic field
domains leading to Eq. \ref{prob-1}. This is however an
  unrealistic oversimplification. More realistically we should expect the
  primordial magnetic fields to be Gaussian random vector fields with a
  power spectrum that depends on the production mechanism \cite{primordialB}. In this case we
  cannot separate the voids into homogeneous regions across which the magnetic fields change abruptly. The variation in magnetic fields across the void would be
  smooth and the adiabaticity parameter, Eq. \ref{Eq:gad} plays an important
  role in this case. For the adiabatic evolution Eq. \ref{Eq:ad} applies and
  the conversion probability would be determined by the high density
  regions at the edge of void with small mixing angles  rather the low density regions near the
center of the void with large mixing angles. For adiabatic evolution therefore we expect the
  photon-axion conversion to be highly suppressed. 

For low axion masses, $\Da \ll \De$ and small mixing angle $\Daga \ll \De$
we have $\cos 2\theta \approx 1$ and the expression for adiabaticity
parameter can be simplified, taking a single Fourier mode $k_B,k_e$ for the
magnetic field and electron distribution respectively,
\begin{align}
\gammaad &\approx  
\left|\frac{\Dosc}{\sin(2\theta)\left(k_B+k_e\right)}\right|\nonumber\\
&\approx 2\left(\frac{\Ne}{10^{-9}~{\rm
      cm^{-3}}}\right)^2\left(\frac{10^{-10}~{\rm
      GeV}^{-1}}{g_{\gammaa}}\right)\left(\frac{1~{\rm
      nG}}{B_T}\right)\left(\frac{0.1 ~{\rm pc}^{-1}}{k_B}\right)\label{Eq:gad2},
\end{align}
where we have assumed that the magnetic field changes more rapidly than the
electron density.
We also need the magnetic field to change randomly in order for Eq. \ref{prob-1}
to be applicable, therefore the evolution should be non-adiabatic w.r.t the
changes in the magnetic field. 
We  {see from Eq. \ref{Eq:gad2} that we} have adiabatic evolution on scales $\gg ~{\rm pc}$, in
particular for Mpc scale magnetic fields considered by
\cite{axion_2}  {rendering their calculations unrealistic}. Most
of the contribution to $\Pga$ would come from magnetic fields on parsec
scales or smaller where the contributions from different domains can add
incoherently. We are therefore in the regime where $\Dosc s\ll 1$ in each
domain of size $s\sim 10$ pc. In this limit we get from Eqs. \ref{prob} and
\ref{prob-1} for a void of radius $\Rv$
\begin{align}
\bar{P}(\gamma \rightarrow a)&\approx \frac{\Pga \Rv}{2 s}\nonumber\\
&\approx \frac{\Daga^2\Rv s }{2}= 10^{-4}\left(\frac{g_{\gammaa}}{10^{-10}~{\rm
      GeV}^{-1}}\right)^2\left(\frac{B_T}{1~{\rm
      nG}}\right)^2\left(\frac{\Rv}{1~{\rm Gpc}}\right)\left(\frac{s}{10~{\rm pc}}\right)\label{Eq:voidlimit}
\end{align}
We note that in this limit the conversion probability is independent of
frequency. 
The COBE-FIRAS limit on change in the CMB frequency spectrum at
the peak of blackbody \cite{firas,fm2002} of
$\lesssim 5\times 10^{-5}$ translates into $g_{\gammaa}B_T \lesssim 10^{-10}~{\rm
  GeV^{-1} nG}$, which is a factor of $\sim 20$ weaker than the limits
obtained in \cite{axion_2}. Our limit is still a very rough limit. To get
precise constraints we must evolve the CMB photons through a realistic void
profile with a realization of Gaussian random magnetic field which we leave for future
work. We can however make the following important observations from above
discussion. We see from Eq. \ref{Eq:gad2} that smaller scales are more
non-adiabatic and should therefore contribute the most to the photon-axion
conversion. However from Eq. \ref{Eq:voidlimit} we see that the conversion
probability decreases for small scales. The conversion probability in a
domain and the effect of adiabaticity therefore oppose each other. We therefore 
have a sweet spot or 
 a range of scales (around $10~{\rm pc}$ for $g_{\gammaa}B_T=10^{-10}~{\rm
   GeV^{-1}nG}$) for which the net conversion probability is maximized. 

\color{black}
\section{ Photon-ALP conversion in the Milky Way halo: Resonant
case}\label{spectral-dist-reso}
\begin{figure}[h]
\centering
\begin{subfigure}[b]{\textwidth}
     \includegraphics[trim={0 0 0 0.8cm}, clip, width=\textwidth]{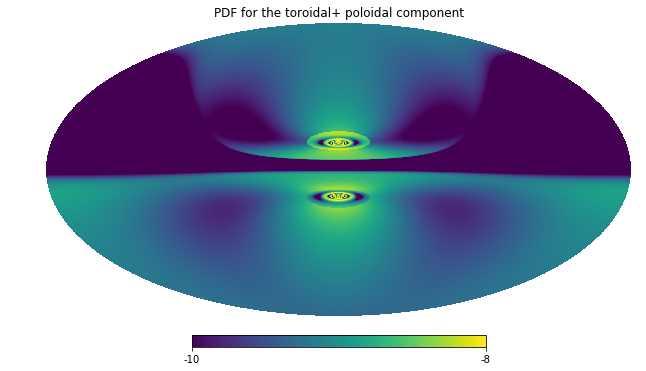}
     \caption{For axion mass $m_a= 5\, \times\, 10^{-12}$ eV}
     \end{subfigure}
  \begin{subfigure}[b]{\textwidth}
     \includegraphics[trim={0 0 0 0.8cm}, clip, width=\textwidth]{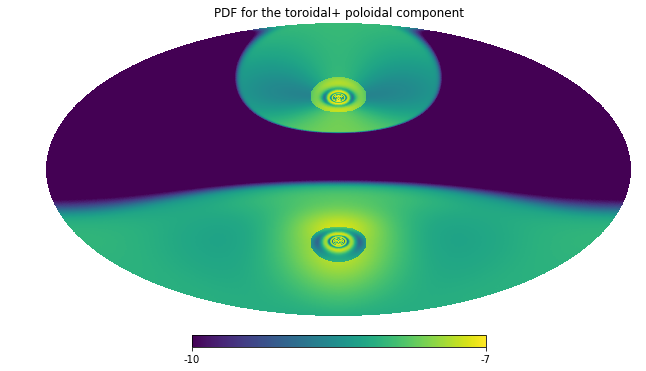}
     \caption{For axion mass $m_a= 5\, \times\, 10^{-13}$ eV}
     \end{subfigure}   
\captionsetup{singlelinecheck=on,justification=raggedright}
\caption{ {Maps} of the resonance conversion signal from photon-ALP at photon
  frequency $\nu=150$ GHz for $g_{11}= 10^{-11}$ GeV$^{-1}$ with Galactic
  magnetic field and electron density is depicted in $\log_{10}$ scale in
  the Galactic coordinates with nside=512 using HEALPix subroutine
  \cite{Gorski:2004by}. The signal  $\Delta I_\nu/I_\nu \geq
    10^{-10}$  corresponds to a sky fraction of (a) $f_{sky}=0.68$ and the
  mean signal of $8.34\, \times \,10^{-10}$ and (b) $f_{sky}=0.41$ and the
  mean signal $1.06\, \times \,10^{-8}$ for the two cases.  }\label{Fig:prob-map-reso}
\end{figure}
The previous work on CMB-ALP conversion only considered the cosmological
evolution of the mean electron density and primordial magnetic fields.  We
can extend the analysis by considering the propagation of CMB photons in
the local Universe through the Milky Way halo to the Solar System. On
scales greater than $\sim 100~{\rm Mpc}$, we expect to approach the
homogeneous Universe \cite{guzzo1997,pc2000,yadav2005,hogg2005,sdb2012,sp2016} and the electron density to
reach the cosmic mean value of $\approx 2\times 10^{-7}~{\rm cm^{-3}}$. The
electron density should increase to  {the Milky Way circumgalactic} value of $\approx
10^{-5}{\rm cm^{-3}}$ near the Miky Way halo at radius of about $\approx
100~{\rm kpc}$ \cite{Miller2013} to $\sim 10^{-1}{\rm cm^{-3}}$ near the
plane of the Galaxy \cite{Cordes:2002wz} before decreasing to $\sim 5\times
10^{-3}{\rm cm^{-3}}$ in the local hot bubble surrounding the Solar system
\cite{cr1987,frs2011,scc2014}. Of course, we do not expect the density to
vary smoothly from  {intergalactic} medium to us but expect the distribution
of matter to be fractal and filamentary
\cite{mc1994,borgani1995,cappi1998,zb2008,ts2008}  and we leave a more careful and
precise analysis taking into account the inhomogeneities in the electron
density for future work. 
We can still get rough constraints using the average electron density
variation from the  {intergalactic} medium to the Solar system. At
$\Ne=\{2\times 10^{-7},10^{-5},10^{-1}\}~{\rm cm^{-3}}$, $\mg=\{1.7\times
  10^{-14},1.2\times 10^{-13},1.2\times 10^{-11}\}~{\rm eV}$
respectively. For $10^{-14}\lesssim \ma\lesssim 10^{-12}$ (the upper limit
coming from the density in the local bubble of $\Ne \sim 5\times 10^{-3}~{\rm cm^{-3}}$), there is only
one resonance and  we can assume production in vaccum
($\cos\theta_0=1$)
and detection at high densities ($\cos\theta=-1$) giving the level
crossing probability (Eq. \ref{Eq:level}) 
\begin{align}
P(\gamma \rightarrow a)&=1-p
\approx \frac{\pi\gammaad}{2}
\approx \frac{2\pi \Daga^2}{|\nabla \De|}
 \lesssim 10^{-4}\label{Eq.reso}\\
\Daga &\lesssim \left(\frac{10^{-4} \left| \nabla \De \right|}{2\pi}\right)^{1/2},
\end{align}\label{Eq.reso1}
where we assumed that $p\approx 1$ to satisfy COBE constraint
 \cite{firas,fm2002} that the fractional  change in CMB spectrum  should be $\lesssim
10^{-4}$. 
For $10^{-14}\lesssim \ma \lesssim 10^{-13}~{\rm eV}$, the resonance
happens outside the Galactic halo with $|\nabla \De| \approx 2.5\times
10^{4} ~{\rm Mpc^{-2}}$ at $100~{\rm GHz}$ and for  $10^{-13}\lesssim \ma \lesssim 10^{-11}~{\rm eV}$ there
will be a resonance inside the Galactic halo with $|\nabla \De| \approx 2.6\times
10^{11} ~{\rm Mpc^{-2}}$ at $100~{\rm GHz}$ giving 
\begin{align}
g_{\gammaa}\BT < 4\times 10^{-10}~{\rm
  GeV^{-1}nG}~ & ~\left| ~10^{-14}\lesssim \ma \lesssim 10^{-13}~{\rm eV}\right.\nonumber\\
g_{\gammaa}\BT < 13\times 10^{-10}~{\rm
  GeV^{-1}\mu G}~ & ~\left|~ 10^{-13}\lesssim \ma \lesssim 10^{-11}~{\rm eV}\right.
\end{align}
We should emphasize an important difference between the last constraint and
the constraints we get on cosmological scales: We know that the Galactic
magnetic field with strength of $\mu {\rm G}$ exists
\cite{jansson,Jansson:2012rt}. The above constraints is therefore directly
on the coupling constant $g_{\gammaa}$ where as the constraints of
\cite{mrs2009} are on the combination $g_{\gammaa}\BT$. 

For $10^{-12}\lesssim \ma\lesssim
10^{-11}$ there will be a second resonance as the photons propagate from
ISM to the local hot bubble surrounding the solar system which would be
less adiabatic and hence give weaker constraints.

For the resonant conversion in the Galactic halo we can use the
  model of Galactic magnetic field and electron number density derived from
 astrophysical  observations including synchrotron radiation, Faraday rotation,
 dispersion of pulsar radiation, angular broadening of extragalactic
 sources and other effects associated with scattering of radiation by
 electrons \cite{jansson,Jansson:2012rt,Cordes:2002wz,Gaensler:2008ec}. The
 details of the Galactic model are given in Appendix
 \ref{sec:galmodel}. Given a model of Galactic magnetic field and electron
 distribution, we can calculate the photon-to axion resonant conversion
 probability for a given axion mass $\ma$ along any direction using Eq. \ref{Eq.reso} at the distance
 $r$ from us where $\ma=\mg$. The results are shown in Fig. \ref{Fig:prob-map-reso}
 for $\ma=5\times 10^{-12},5\times 10^{-13}~{\rm eV}$ assuming
 $g_{\gammaa}=10^{-11}~{\rm GeV}^{-1}$ at observed frequency $\nu=150~{\rm GHz}$. A given axion mass traces a
 complicated shell around the Galaxy where $\ma=\mg$ resulting in rich
 features in the CMB spectral distortion map. In particular, given a
 Galactic model,  each mass $\ma$ has its unique morphological signature in
 the CMB sky which is quite different from any known cosmic or Galactic
 foregrounds and backgrounds. The north-south
  asymmetry in Fig. \ref{Fig:prob-map-reso} is a reflection of the north-south asymmetry in the Galactic
  magnetic field (see Appendix \ref{sec:galmodel}). We should emphasized that in addition 
 this distortion is $100\%$ polarized and therefore can be easily
 distinguished from non-polarized cosmic and Galactic components. The lower
 axion masses come into resonance further out in the halo where the
 electron number density is smaller giving higher distortions.

\subsection{Distinguishing between scalars and pseudoscalars using polarization}\label{sec:lsp}
{If we had new low mass scalar particles (LSPs) mixing with the photons \cite{raffeltbook}, we would
  get a similar anisotropic distortion pattern as in the case of
  pseudoscalars such as ALPs. The interaction for scalars ($\phi$) is given by 
\begin{align}
\mathcal{L}_{\rm int}=g_{\gamma
    \phi}\mathbf{B_{\gamma}.B_{\rm ext}} {\phi}\label{Eq:scalar}
\end{align}
and should be compared to Eq. \ref{Eq:int}. In Eq. \ref{Eq:scalar}
$\mathbf{B}_{\gamma}$ is the magnetic field of the photon and $\mathbf{B}_{\rm ext}$ is the
external magnetic field.
For the scalars therefore, in the presence of external magnetic field, the
photon polarization with its magnetic field along the external magnetic
field is coupled to the axions and therefore the polarization of the
distortion is rotated by $90^{\circ}$ compared to the pseudoscalars or ALPs. For equivalent couplings, we will therefore
have the same anisotropy signal on the sky but orthogonal polarization. The
polarized signal discussed in this section, if detected, will not only tell
us whether there is a light particle that mixes with photons and the mass
of this particle but also
whether it is a scalar or a pseudoscalar.
}

\section{Photon-ALP conversion in the Milky Way halo: Non-resonant case}\label{non-resonant}
There is no resonance for
$\ma \lesssim  10^{-14}~{\rm eV}$ except for the 
resonances expected when neutral gas is encountered with very low
ionization fraction \cite{mrs2009} and we will  ignore these as they
require a detailed multiphase model of the Galaxy.  For
$\ma > 10^{-11}\, \text{eV}^{-1}$, the required values of $g_{\gammaa}$ to
produce any observable spectral distortions are ruled out by CAST
\cite{cast}. We will consider non-resonant conversion  for $\ma \lesssim  10^{-14}\,
\text{eV}$ in this section.

For Milky way halo,
the typical length of the coherent magnetic field is of kpc scale with the strength of $\mu G$ \cite{jansson,Jansson:2012rt}. In addition,
there is a turbulent component to the magnetic field in the Milky Way confined mostly in the
  Galactic plane with
  coherent lengths of 100 pc or less
  \cite{bkb1985,srwe2008,hbg2008}. Galactic magnetic field and electron
  density distribution in the Galactic halo is not yet known very
  accurately, but the situation is expected to improve with the upcoming
  missions like SKA \cite{2015aska.confE..41H} in the future.
 The recent measurement of the Galactic magnetic field
from Faraday rotation map \cite{Oppermann:2011td} and Planck dust map
\cite{2016A&A...596A.103P} also indicates fluctuations in the magnetic
field dominant at large angular scales. As a result, we can also expect
large-scale fluctuations in the component of Galactic magnetic field
transverse to the direction of propagation. For the purpose of
  photon-axion conversion, what is important is not the fluctuations in
  electron density but rather the fluctuations in photon effective mass
  which can transition from real to being imaginary in the highly neutral
  gas. Fluctuations in the electron ionization fraction can thus make the
  propagation of photons/axions highly non-adiabatic and should be
  important for non-resonant conversion.

\subsection{Coherent magnetic fields and electron distribution}

A model of the coherent component of the Galactic magnetic field
  in the disk and halo of Milky way was developed by Jansson et
  al. \cite{jansson}. The Magnetic field in the Galactic halo can be
  written as a superposition of a toroidal and poloidal components. In both
these components the magnetic field changes on scales of $\sim ~{\rm
  kpc}$ (see Appendix \ref{sec:galmodel} for details).
\begin{figure}[H]
\centering
     \includegraphics[trim={0 0 0 0.8cm}, clip, width=\textwidth]{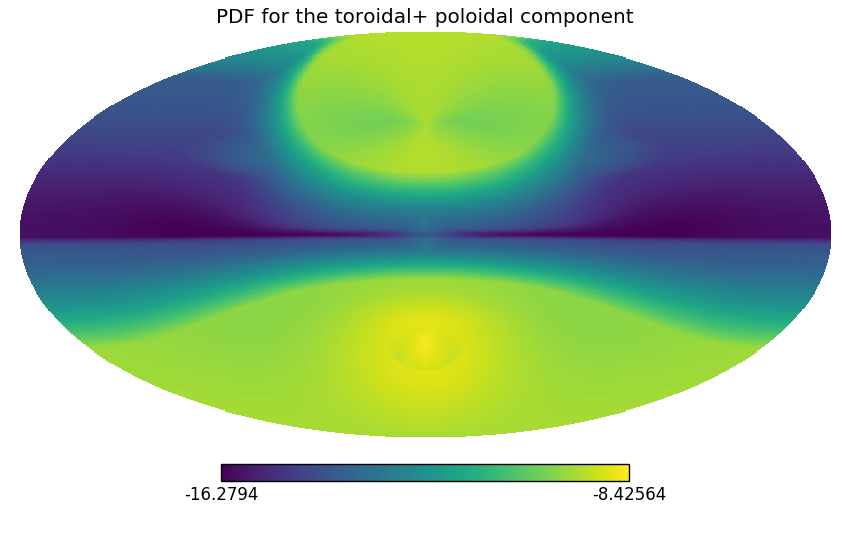}
\captionsetup{singlelinecheck=on,justification=raggedright}
\caption{A map of the maximum probability of conversion from photon-ALP at photon frequency $\nu=500$ GHz in the Galactic coordinates with nside=1024 using HEALPix subroutine \cite{Gorski:2004by} are depicted in $\log_{10}$ scale }\label{Fig:prob-map}
\end{figure}

The electron density also decreases exponentially
 with increasing distance from the Galactic plane
 \cite{Cordes:2002wz,Gaensler:2008ec,Miller2013} with a scale height again of
 order $\sim ~{\rm kpc}$. 
The photon-ALPs conversion probability, Eq. \ref{prob},  is proportional to $\BT^2/\Ne^2$
 for $\ma \lesssim 10^{-14}~{\rm eV}$. Since both $\BT$ and $\Ne$ decrease with increasing distance from the  Galactic center and the Galactic
 plane, there will be a maximum conversion probability at some distance $s$
 for each direction in the sky and hence we can have a map of this
 effect. This map provides an upper limit to the axion spectral
   distortion we can expect in the CMB. Using the model of the Galactic magnetic field and electron density mentioned above, we obtain map of the maximum photon-ALPs conversion probability $P (\gamma \rightarrow a) (\hat n)$ as a function of the direction of the sky in Fig. \ref{Fig:prob-map} using HEALPix \cite{Gorski:2004by} with nside=1024. 
 \begin{figure}
\centering
 \includegraphics[width=\textwidth]{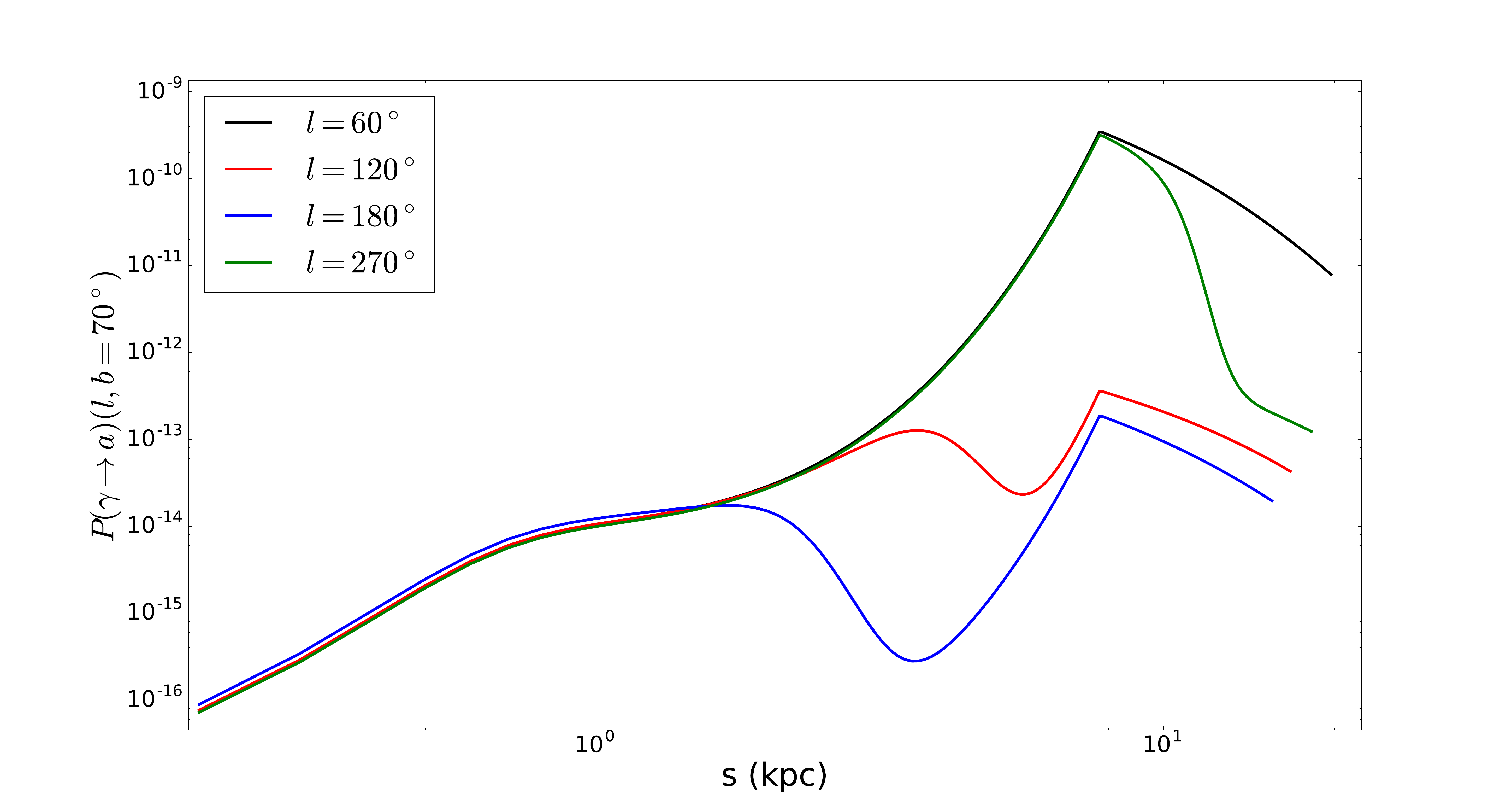}
\caption{Probability of photon-ALP conversion at photon frequency
    $\nu=500$ GHz as a function of distance $s$ from Sun for different sky
    directions  with the large scale  Galactic magnetic field from Jansson
    et al. \cite{jansson} and the electron density model from Cordes et
    al. \cite{Cordes:2002wz} with the revised parameters according to
    Gaensler et al. \cite{Gaensler:2008ec}. We assumed that the outer
    distribution of electron goes to a constant value \cite{Miller2013}
    giving the sharp change in curve at large distances.}\label{Fig:bne}
\end{figure}

From Eq. \ref{Eq:gad2} we see that for the Galactic parameters we will have
adiabatic evolution for scales $\gtrsim 10^{-4}~ {\rm pc}$, 
\begin{align}
\gammaad 
&\approx 2\left(\frac{\Ne}{10^{-5}~{\rm
      cm^{-3}}}\right)^2\left(\frac{10^{-10}~{\rm
      GeV}^{-1}}{g_{\gammaa}}\right)\left(\frac{1~{\rm
      \mu G}}{B_T}\right)\left(\frac{10^4 ~{\rm pc}^{-1}}{k_B}\right)\label{Eq:gad3}
\end{align}
a condition
easily satisfied by the large scale average magnetic field and electron
distribution. For adiabatic evolution (Eq. \ref{Eq:ad})  the final conversion probability in the
  large scale coherent magnetic field of the Galaxy would be decided by the
  mixing angles very close to us in the high density region of the Galaxy. In Fig. \ref{Fig:bne} we show the evolution of the $\Pga$
from outskirts of the Galactic halo to solar neighborhood. The final
conversion probability is given by the value near the observer which is
many orders of magnitude smaller than the maximum value at around 8 kpc
(see also Fig. \ref{Fig:prob-map}). Thus for all practical purposes we CMB spectral
distortion contribution from the large scale morphology of the Galactic
magnetic field can be neglected if we do not take into account the
turbulence in the interstellar medium.

\subsection{Random magnetic field and turbulent gas}

We can see from Eq. \ref{Eq:gad} that there are two ways to avoid adiabatic
evolution: large gradients in either the magnetic field $B_T$ or photon effective
mass $\meff$. There is observational evidence for turbulence in the interstellar
electrons from 100 pc scales down to sub-parsec scales \cite{ars1995} with
Kolmogorov like power law. If the source of gas turbulence is
magneto-hydrodynamics (MHD), we should expect stochastic magnetic fields on
sub-parsec scales also. For turbulence on scales $s=10^{-4}~{\rm pc}$ in
regions of size $R\sim 1000$ pc we get
\begin{align}
\bar{P}(\gamma \rightarrow a)&\approx \frac{\Pga R}{2 s}\nonumber\\
&\approx \frac{\Daga^2 R s }{2}= 10^{-9}\left(\frac{g_{\gammaa}}{10^{-10}~{\rm
      GeV}^{-1}}\right)^2\left(\frac{B_T}{1~{\rm
      \mu G}}\right)^2\left(\frac{R}{~{\rm 1000 pc}}\right)\left(\frac{s}{10^{-4}~{\rm pc}}\right)\label{Eq:ismlimit}
\end{align}
These distortions as well as the cosmological distortions from random
magnetic fields in voids would be unpolarized.

\section{Forecasts for CORE, LiteBIRD and PIXIE}\label{forecast} 
The detectability of the temperature and polarization spectral distortions
in the CMB would depend on sensitivity as well as frequency coverage and
number of channels of a CMB experiment. The frequency coverage
(i.e. channels covering the full CMB spectrum from Rayleigh-Jeans to Wien
region)  and
sufficient number of frequency channels are essential if we are to
distinguish between the axion spectral distortions, primary CMB
anisotropies, $y$-type and $\mu$-type  distortions and Galactic and
extragalactic foregrounds. In the case of anisotropic axion distortions coming
from the photon-ALP conversion in the Galactic magnetic field, we can also
use the morphology of the predicted signal to distinguish it from other components.

To estimate the detectability of the signal from CMB missions
  using the spectrum information alone, we do a
Fisher matrix analysis \cite{Tegmark:1996bz}. In reality we will also have
spatial anisotropy information from the Galactic model or looking towards
known voids. Our estimates from Fisher analysis should therefore be
considered conservative.  We  model  the
observed intensity difference from the Planck spectrum with temperature
$T_{\rm CMB}=2.7255~{\rm K}$,  $\Delta I_{\nu}$, as 
\begin{align}\label{fisher_b}
\begin{split}
\Delta I_{\nu}=  \Delta T_{\rm CMB} s^{\text{CMB}}_{\nu} + A_{\gammaa}
s^{\gammaa}_{\nu} + y s^{\text{y}}_{\nu} +  {A_{\rm sync} s^{\text{sync}}_{\nu} +} A_{\text{Dust}}s^{\text{Dust}}_{\nu},  
\end{split}
\end{align}
here we have defined \cite{Adam:2015wua, 1998ApJ...505..473P}
\begin{align}\label{fisher_c}
\begin{split}
& {s^{\text{sync}}_{\nu} = \bigg(\frac{2 k_{\rm
    B}\nu^2}{c^2}\bigg)\bigg(\frac{\nu_s}{\nu}\bigg)^\alpha,\,\, \nu_s= 30\, \text{GHz},}\\
&s^{\text{CMB}}_{\nu} = \frac{2 \kB\nu^2}{c^2}\frac{x^2
  e^x}{(e^x-1)^2}, \hspace{3cm} x= h\nu/(\kB T_{\text{CMB}}),\, T_{\text{CMB}}= 2.7255\, \text{K},\\
&s^{\text{Dust}}_{\nu}= \frac{2 k_{\rm
    B}\nu^2}{c^2}\bigg(\frac{\nu}{\nu_0}\bigg)^{\beta_d+1}\bigg(\frac{e^{h\nu_0/(\kB T_d)}-1}{e^{h \nu/(\kB T_d)}-1}\bigg), \hspace{0.5cm} T_d=18 \text{K}, \nu_0= 545 \text{GHz},\\
&s^{\text{y}}_{\nu}= \frac{2h\nu^3}{c^2}\bigg(\frac{xe^x}{(e^x-1)^2}\bigg)\bigg(\frac{x(e^x+1)}{e^x-1}-4\bigg),
\end{split}
\end{align}
where $\Delta T_{\rm CMB}$ is the CMB temperature anisotropy in CMB
  temperature units of
  $K_{\rm CMB}$, $A_{\rm Dust}$ is the brightness temperature of dust at
  $\nu_0=545~{\rm  GHz}$ and ${y}$ is the dimensionless amplitude of the
  $y$-type distortion or the thermal Sunyaev-Zeldovich (tSZ) effect.

For the resonant conversion case, we can write for the $100\%$
  polarized signal, 
\begin{align}
s^{\gammaa}_{\nu}= \frac{h\nu^3}{c^2}\left(\frac{\nu}{\nu_0}\right)\frac{\mathcal{I}^{\gammaa}(\nu_0,\ma)}{(e^x-1)},
\end{align}
where $-\mathcal{I}^{\gammaa}(\nu_0,\ma)$ is the probability of conversion
at $\nu_0=150~{\rm GHz}$ for axion mass $\ma$ for the fiducial Galactic
model (see Eq. \ref{Eq.reso}) with
coupling $g_{\gammaa}=10^{-10}~{\rm GeV}^{-1}$ and the dimensionless amplitude is defined
as
\begin{align}
A_{\gammaa}\equiv \left(\frac{g_{\gammaa}}{10^{-10}~{\rm GeV}^{-1}}\right)^2
\end{align}
This polarized distortion is not degenerate with the Sunyaev-Zeldovich
effect or the $y$-type distortion \cite{zeldovich} and we can ignore the $y$ component
while fitting for it. The polarization pattern of this distortion will also
be very different from the CMB primary and lensing polarization. In
particular polarized axion distortion  will have both E and B modes. We will therefore
assume that the morphological information will separate the polarized axion
distortion from the CMB primary and lensing polarization signals. The only
serious contamination is therefore expected from the Galactic dust emission
and also synchrotron emission if the low frequency channels (below 100 GHz)
are also used.  {We use the complete frequency range and include synchrotron as well as dust contaminations in the Fisher analysis. We also do an analysis with only dust and frequency channels higher than $100$ GHz and compare the Fisher matrix forecasts in Table \ref{tab_2}. We see that the presence of synchrotron radiation degrades constraints by a factor of $\sim 2$.  {Future experiments like CBASS \cite{2015MNRAS.448.3572I} and NEXTBASS \footnote{\url{https://www2.physics.ox.ac.uk/research/experimental-radio-cosmology/the-next-band-all-sky-survey-nextbass}} can make improvements in understanding the synchrotron emissions at low frequency. Use of these experiments jointly with LiteBIRD and CORE will improve measurability of the signal. By using the unique spatial structure of the photon-axion conversion signal, one can perform a spatial template base search in the data. This will enable further improvements in the Fisher estimates. So, our estimates presented here are very conservative and expected to improve in the future.}}

For the non-resonant conversion we have,
\begin{align}
s^{\gammaa}_{\nu}= \left(\frac{2h\nu^3}{c^2}\right)\frac{\mathcal{I}^{\gammaa}(R,s)}{(e^x-1)},
\end{align}
where $-\mathcal{I}(R,s)$ is the frequency independent probability of
conversion of unpolarized intensity (Eqns. \ref{prob-1},\ref{Eq:voidlimit},\ref{Eq:ismlimit})
for turbulent magnetic fields of coherence length $s$ for photons
traversing a distance $R$ for coupling $g_{\gammaa}=10^{-10}~{\rm GeV}^{-1}$
 and the dimensionless amplitude in this case is defined
as
\begin{align}
A_{\gammaa}&\equiv \left(\frac{g_{\gammaa}B_T^{\rm rms}}{10^{-10}~{\rm
      GeV}^{-1}~{\rm nG}}\right)^2& :{\rm voids},\label{Eq:defAgturbvoid}\\
A_{\gammaa}&\equiv \left(\frac{g_{\gammaa}B_T^{\rm rms}}{10^{-10}~{\rm
      GeV}^{-1}~{\rm \mu G}}\right)^2& :{\rm Galaxy},\label{Eq:defAgturb}
\end{align}
where $B_T^{\rm rms}$ is the magnetic field strength on scales $s$.
In this case the distortion is unpolarized and we must marginalize over the
$y$-type distortion and CMB anisotropies.

\begin{figure}
    \centering
        \includegraphics[width=\textwidth,keepaspectratio=true]{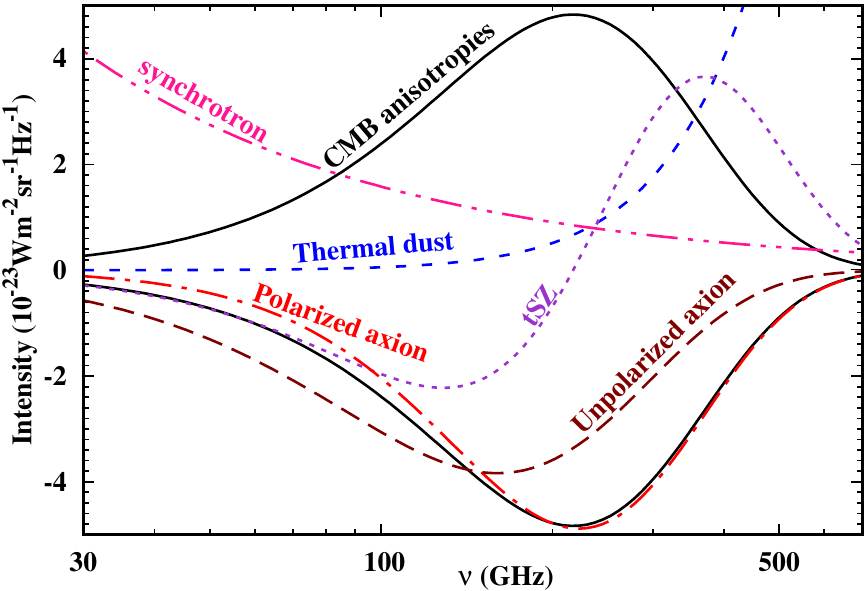}
           \caption{{The spectra of  the components mentioned in
             Eq. \eqref{fisher_b}  are compared with $A_{\rm CMB}=\pm 10~{\rm
               \mu K}$ for the CMB anisotropies, $A_{\rm dust}=1~{\rm \mu
               K}$ at 545 GHz for thermal dust emission, $A_y=2\times 10^{-6}$ for
             thermal Sunyaev-Zeldovich (tSZ) effect,   {$\beta_d = 1.4$ \cite{Adam:2015wua}, for synchrotron $A_\text{sync}=150\, \mu K$ at 30 GHz, $\alpha=2.8$ \cite{1998ApJ...505..473P}}, 
             $A_{\gammaa}=20, \bar{\mathcal{ I}}(\ma=5\times 10^{-13}~{\rm
               eV})=-10^{-6}$ for polarized axion distortion and 
             $A_{\gammaa}\bar{\mathcal{I}}=-10^{-5}$ for the unpolarized
             axion distortion.} \label{spectrum-1}}
 \end{figure}

The spectrum of each of the signal is plotted in Fig. \ref{spectrum-1} with
the amplitudes chosen so that the intensities are of similar amplitude
allowing us to compare the shapes of the spectra.
The detectability of photon-ALP conversion depends upon the error budget
of a particular mission, its frequency coverage and
  number of frequency channels available.  The measurability of a non-degenerate parameter $(p_i)$ depends
upon the covariance matrix $(\mathcal{C}_{i \,i})$,  which in turn depends
upon the inverse of the Fisher matrix elements ($\mathcal{C}_{i
  \,i}=1/F_{ii}$) \cite{Tegmark:1996bz}, The elements of Fisher matrix for
a set of parameters $\mathcal{P}= ( p_1,\, p_2,\, \hdots,\, p_n)$ are
defined as \cite{Tegmark:1996bz} 

\begin{align}\label{fisher_d}
\begin{split}
F_{ij}= \sum^n_{\alpha=1} \frac{\partial \Delta s(\nu_\alpha)}{\partial p_i}\frac{1}{(\Delta s^{n}_{\nu})^2}\frac{\partial \Delta s(\nu_\alpha)}{\partial p_j},
\end{split}
\end{align}
where, the sum is made over all frequency channels and $\Delta s^{n}_{\nu}$
denotes the value of instrumental noise specific to a particular
mission. For the degenerate parameters the Fisher matrix is not diagonal.
 The elements of covariance matrix in this case are computed from the
inverse of the Fisher Matrix  ($\mathcal{C}_{ij}=
(F^{-1})_{ij}$).  We use the model given in  Eq. \eqref{fisher_b},
  with the parameter vector for resonant polarized distortion given by
  $\mathcal{P}= (  A_{\gammaa},\, A_{\text{Dust}},\,
  \beta_d,  {\alpha,\, A_{sync}})$ and for non-resonant unpolarized distortion given by
  $\mathcal{P}= ( \Delta T_{\text{CMB}},\, A_{\gammaa},\, A_{\text{Dust}},\, \beta_d,\, y\,,  {\alpha,\,  A_{sync}})$. We calculate the Fisher matrix at following fiducial values of foreground model,  {$A_d= 100\, \mu$K at $545$ GHz, $\beta_d= 1.5 , A_{\text{sync}}= 150\, \mu$K at $30$ GHz, $\alpha= 2.8$ \cite{Adam:2015wua}.}
  
\begin{table*}[h]
\centering
\caption{Instrumental noise for different missions}
\label{tab_1} 
\vspace{0.5cm}
\begin{tabular}{|p{1.5cm}|p{2.5cm}|p{4.0cm}|p{4.0cm}|p{1.5cm}|}
\hline 
\centering Mission& \centering Frequency Channels (GHz) & \centering
Instrumental noise & \centering All sky sensitivity \\\centering ($ 10^{-27} {\rm Wm^{-2}sr^{-1}Hz^{-1}}$) & \centering Duration (Months)  \tabularnewline
\hline
PIXIE & \centering  {30}-600 $\Delta \nu=$15  & \centering ($\Delta
I^I_\nu=4 \times 10^{-24}$ \& \\  $\Delta I^{P}_\nu=6 \times 10^{-25}$)\\
${\rm W m^{-2}Hz^{-1} sr^{-1}}$ per pixel. \\Total number of Pixels (N)$= 49152$ & \centering $\Delta I^I_\nu=18 $  \\  {$\Delta I^{P}_\nu=2.7$} &  \centering 48 \tabularnewline
\hline
LiteBIRD & \centering  {40, 50, 60, 68, 78, 89,} 100, 119, 140, 166, 195, 235, 280, 338, 402  &
\centering $w_T^{-1/2}=$ ( {26, 16.7, 13.8, 11.2, 9.4, 8.1,} 6.4, 5.3, 4.1, 4.5, 4.0, 5.3, 9.2, 13.5, 26.1)
$\mu $K arcmin\\ $ w_P^{-1/2}=$ ( {36.8, 23.6, 19.5, 15.9, 13.3, 11.5,} 9.0, 7.5, 5.8, 6.3, 5.7, 7.5, 13.0, 19.1,
36.9) $\mu$K arcmin &  \centering $\Delta I^I_\nu =$ ( {1, 0.98, 1.14, 1.15, 1.23, 1.32, }1.25, 1.3, 1.25, 1.6,
1.55, 2.1, 3.2, 3.6, 4.5)  \\ {$ \Delta I^P_\nu =$ }( {1.42, 1.4, 1.6, 1.64, 1.74, 1.88, }1.8, 1.9, 1.8, 2.2, 2.2, 2.9, 4.8, 5.1, 6.3)  &\centering 36\tabularnewline
\hline
CORE & \centering  {60, 70, 80, 90,} 100, 115, 130, 145, 160, 175, 195, 220, 255, 295, 340,
390, 450, 520, 600 & \centering {$w_T^{-1/2}$=} ( {7.5, 7.1, 6.8, 5.1, } 5.0, 5.0, 3.9, 3.6, 3.7, 3.6,
3.5, 3.8, 5.6, 7.4, 11.1, 22.0, 45.9, 116.6, 358.3)\\ $\mu$K arcmin\\ {$ w_P^{-1/2} =$} ( {10.6, 10.0, 9.6, 7.3,} 7.1, 7.0, 5.5, 5.1, 5.2, 5.1, 4.9, 5.4, 7.9, 10.5, 15.7, 31.1, 64.9, 164.8, 506.7 )\\$\mu$K arcmin &  \centering $\Delta I^I_\nu =$ ( {0.62, 0.77, 0.93, 0.84}, 1.0, 1.2, 1.1, 1.1, 1.3, 1.3, 1.4, 1.5, 2.1, 2.5, 2.9, 4.1, 5.3, 7.0, 9.3 ) \\ $\Delta I^P_\nu =$ ( {0.88, 1.09, 1.31, 1.21,}1.4, 1.7, 1.5, 1.6, 1.8, 1.9, 1.9, 2.1, 3.0, 3.5, 4.1, 5.9, 7.5, 9.9, {13}) & \centering 36 \tabularnewline
\hline
\end{tabular}
\end{table*}

\subsection{Future constraints from the unpolarized  axion distortion (non-resonant conversion)} 

\begin{figure}
    \centering
    \begin{subfigure}[b]{0.9\textwidth}
    \centering
        \includegraphics[width=0.9\textwidth]{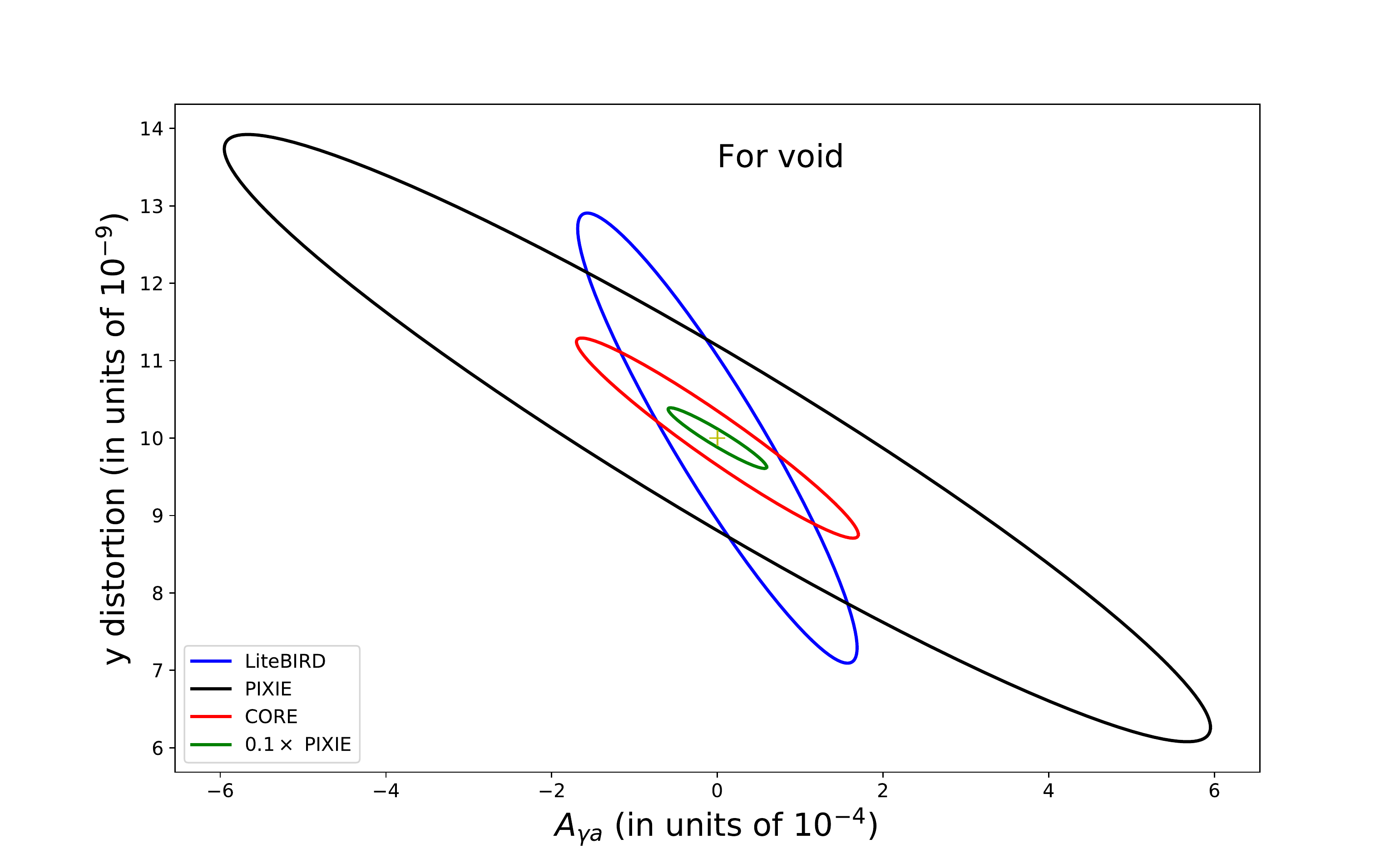}
    \end{subfigure}
    ~ 
    \begin{subfigure}[b]{0.9\textwidth}
    \centering
        \includegraphics[width=0.9\textwidth]{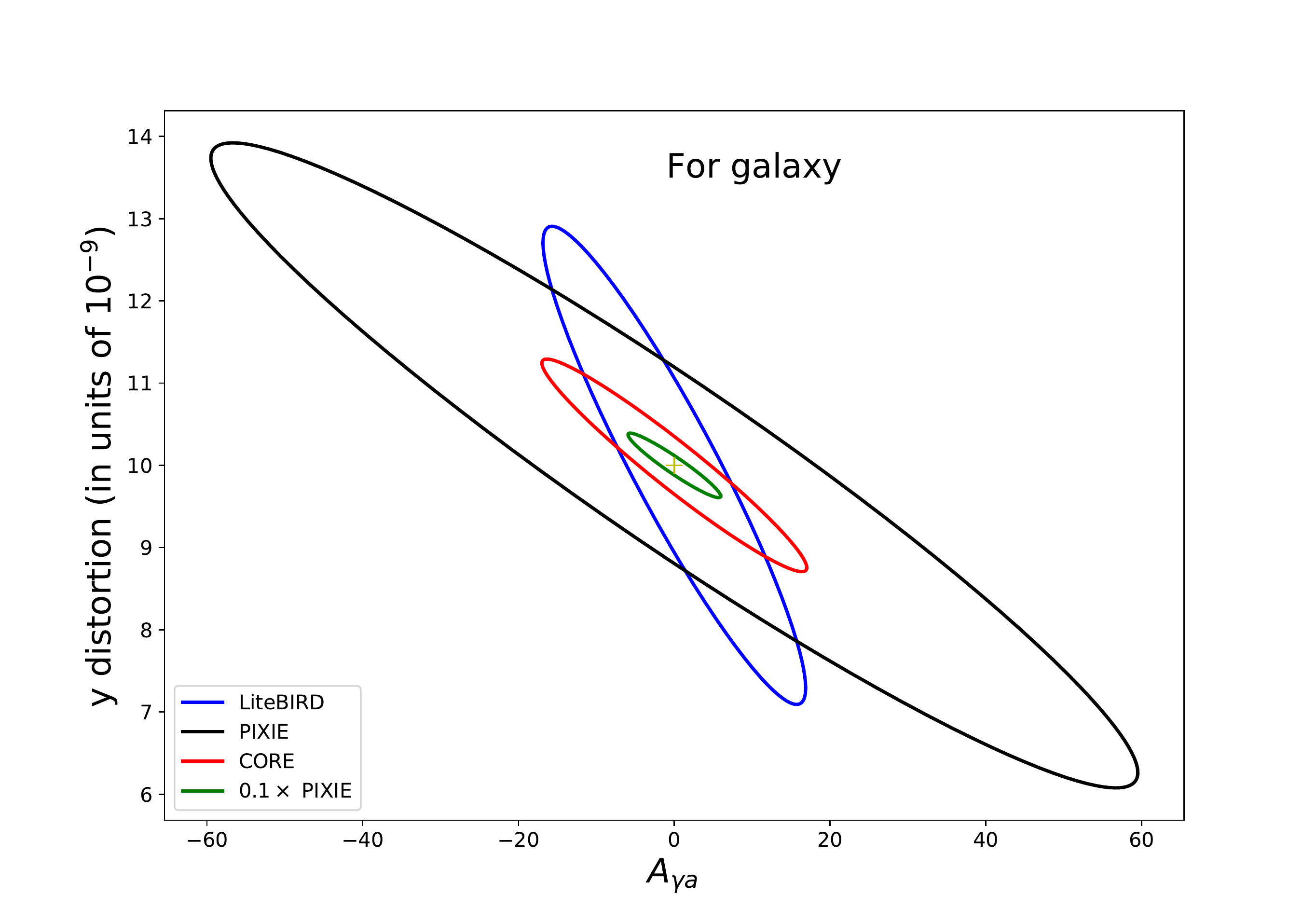}
    \end{subfigure}
       \caption{{$68\%$ contour between $y$ distortion and
           $A_{\gammaa}\propto \left(g_{\gammaa}B_{T}^{\rm rms}\right)^2$ (defined in Eq. \ref{Eq:defAgturb} for the Galaxy and Eq. \ref{Eq:defAgturbvoid}
           for the voids) for
           different future missions.  In green (the smallest contour), we plot for a case with instrumental noise better than PIXIE by a factor of $10$.}}\label{fisher-2}
\end{figure}
 
 { The results of the Fisher analysis are shown in
   Fig. \ref{fisher-2} for future CMB missions like PIXIE \cite{Kogut:2011xw},
 CORE\cite{2017arXiv170604516D} and LiteBIRD  \cite{Matsumura:2016sri} and
 also}  for a mission with $10$ times better {sensitivity} than
PIXIE.   {The Fisher analysis are performed for two scenarios, (i) with only dust as the foreground contaminations and (ii) with both synchrotron and dust contaminations. For case (i), we have used 
 the frequency channels between $100-600$ GHz for PIXIE and CORE
 and $100-402$ GHz for LiteBIRD with the corresponding instrumental noise
 mentioned in Table \ref{tab_1}, $f_{sky}=1$ and using Eq. \ref{fisher_b}
 for the estimation. We have marginalized out CMB \& dust and obtained the
 contour for $y$ distortion and $A_{\gammaa}$ for {photon-axion
   conversion in } the turbulent 
 magnetic field in voids { and in our Galaxy.} For the case (ii), we used all the frequency channels provided in Table \ref{tab_1} and have marginalized over the amplitude as well as the spectral index of synchrotron and dust in the Fisher analysis. The CORE Fisher estimates gets worse by nearly a factor of $\sim\, 5$, after considering both synchrotron and dust due to the absence of a few low frequency channels. For missions like LiteBird and PIXIE, the constraints degrades constraints by a factor of $\sim 2$.  {The Fisher estimates presented here are conservative and can be further improved by using the spatial template of the photon-axion conversion signal.} }

{The signal and thus our constraints depend not only on the
  quantity the coupling $g_{\gammaa}$ and magnetic field strength $B_T^{\rm
    rms}$ which make up the amplitude $A_{\gammaa}$
  (Eq. \ref{Eq:defAgturbvoid}, \ref{Eq:defAgturb}) but also the void or Galaxy model  as shown in Eqs. \eqref{Eq:voidlimit} and
 \eqref{Eq:ismlimit} respectively. We can rescale the constraints for
 different void or Galactic model using Eqns. \ref{Eq:voidlimit} and \ref{Eq:ismlimit}. The
 constraints in Fig. \ref{fisher-2}} for the voids are for the parameters mentioned in
 Eq. \eqref{Eq:voidlimit} with $R_v= 1$ Gpc and $s_v=10$ pc. For the
 Galaxy, {the forecasts} are obtained for $R_g= 1$ kpc and $s_g=10^{-4}$
 pc. Our results show that the turbulent component of the voids can impose
 stronger constraints than the Galaxy. This is
 {because of
 larger scale ($R$) of the voids as well as the larger turbulence scale
 ($s$) when the propagation becomes non-adiabatic compared to the
 Galaxy. The larger non-adiabaticity scale ($s$) is in turn the result of small
 electron densities in the voids (Eq. \ref{Eq:gad2}). We should therefore
 expect strongest constraints from the emptiest voids for the same magnetic
field strength.}   Stacking
 the known voids from other cosmological
 probes could also improve the SNR  and {we leave a detailed study
   with more realistic void profile and magnetic field spectrum for
 future work.}
 
\subsection{Future constraints  from the  polarized anisotropic axion
  distortion (resonant conversion)}
{The $100\%$ polarized anisotropic spectral distortion} from the resonant conversion in
the Galactic magnetic field can evade the contamination from {statistically
isotropic} $y$-distortion
and {CMB ansiotropies} due to its {characteristic
  polarization pattern} in the sky. {We can therefore assume that
  these components would be separated using the morphological and
  polarization information and ignore them for the Fisher analysis. The
  only significant contamination we must distinguish (above 100 GHz) is then the Galactic dust
  contamination.}  In Fig. \ref{fisher-4}, we plot the possible constraints
which can be obtained from the polarized signal in presence of  {synchrotron and} dust after
marginalizing  {over}  {$\alpha,\, A_{\text{sync}},\, \beta_d$}.  
{We use the mean signal calculated in Sec. \ref{spectral-dist-reso} and Fig. \ref{Fig:prob-map-reso} from the parts of
the sky with signal $\Delta I_\nu/I_\nu >10^{-10}$. This selects a fraction
of sky $f_{sky}=0.68$ for axion mass $\ma=5 \times 10^{-12}~{\rm eV}$ with average
distortion $\bar{\mathcal{I}}(\nu=150~{\rm GHz})=8.3\times 10^{-8}$ and
$f_{sky}=0.4$ for axion mass $\ma=5 \times 10^{-13}~{\rm eV}$ with average
distortion in this fraction of sky of $\bar{\mathcal{I}}(\nu=150~{\rm
  GHz})=1.1\times 10^{-6}$ for $g_{\gammaa}=10^{-10}~{\rm GeV}^{-1}$. We do the   Fisher analysis for
the two  axion masses using these mean distortions.  The signal for low
mass axions is
stronger and hence can be better constrained at high latitudes in
comparison to the signal from high mass axions (see Eq. \ref{Eq.reso} and
Sec. \ref{spectral-dist-reso}) because the
resonance happens further out in the Galaxy where the electron density is
smaller.  Our knowledge of electron distribution and magnetic fields in the
Galaxy should improve considerably in not far future, on the similar timescales as the future
CMB missions. We should therefore expect strong 
direct constraints on the  photon-axion coupling from the polarized
anisotropic distortions of the CMB in the future in the axion mass range where
resonances can occur in the Galactic halo.  {A Fisher forecast for different combinations of the foreground is presented in Table \ref{tab_2}. The second column of Table \ref{tab_2} is obtained with only dust as the foreground contaminations and we used frequency channels above $100$ GHz. The Fisher forecasts for the more realistic situation including both synchrotron and dust and using all frequency channels are shown in the second column in Table \ref{tab_2}.
The presence of synchrotron degrades the constraints by little more than a factor of $2$.}

\begin{table*}[t]
\centering
\caption{ {Fisher forecast for two different combinations of foregrounds Dust (D) and Synchrotron (S). For dust (D) only case, we used channels only above $100$ GHz. For dust and synchrotron (S+D) case, we used all the frequency channels shown in Table \ref{tab_1}. The constraints we get for S+D case are around $2$ times weaker compared to the case when only dust foreground is present. For CORE, the constraints degrades by around $5$ times due to the absence of low frequency channels.  {These estimates are conservative and can be improved by using the unique spatial structure of the photon-axion conservation signal.}
}}
\label{tab_2} 
\vspace{0.5cm}
\begin{tabular}{|l|l|l|l|l|l|l|}
   \hline
    \multirow{2}{*}{Probe} &
      \multicolumn{3}{c|}{$[F^{-1}]_{ii}$  with D  {only} } &
      \multicolumn{3}{c|}{$[F^{-1}]_{ii}$ with S+D}\\
    & PIXIE & LiteBIRD & CORE & PIXIE & LiteBIRD & CORE \\
\hline
$g^2_{10}$ \newline for  $m_{a} = 5 \times 10^{-13}$ eV \newline ($\times 10^{-3}$)& \centering $0.62$ &  $1.92$  &  $0.83$ &
  \centering  $1.29$ &  $3.5$ & $2.28$ \\
\hline
$A_{\gamma a}$ for void ($\times 10^{-4}$) &  \centering  1.63 &  $0.48$ &  $0.23$ 
 &\centering  3.9 &  $1.1$  &  $1.12$  \\ 
\hline
$A_{\gamma a}$ for galaxy &  \centering  16.3 &  $4.83$ &  $2.25$ 
 & \centering  39  & $11.1$ &  $11.2 $ \\ 
\hline
\end{tabular}
\end{table*}

\begin{figure}[H]
\centering
\begin{subfigure}[b]{0.9\textwidth}
\centering
\includegraphics[width=0.9\textwidth]{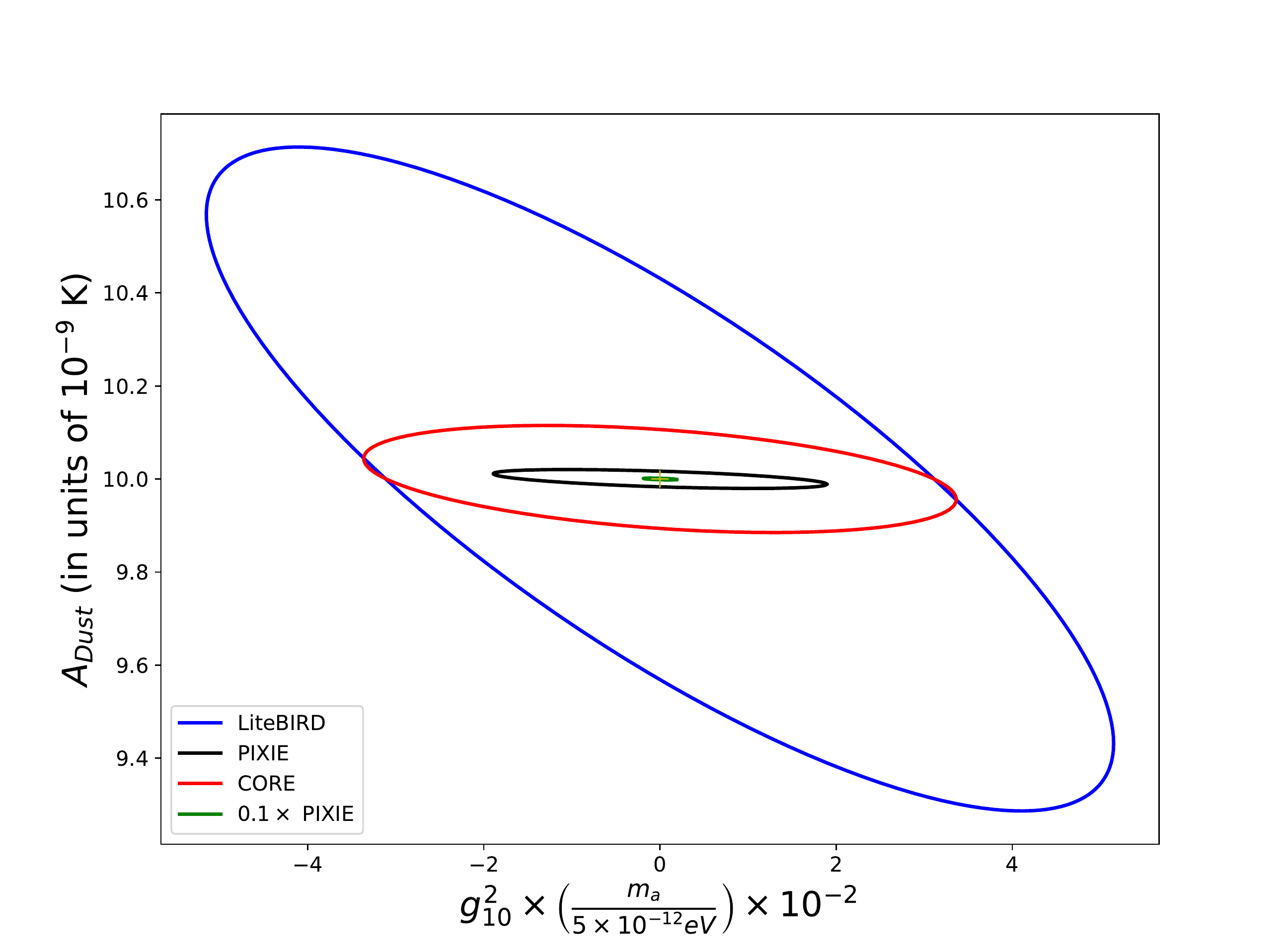}
\captionsetup{singlelinecheck=on,justification=raggedright}
\end{subfigure}
\caption{{$68\%$ contour between dust contamination $A_{Dust}$ and
    $g_{\gammaa}^2\times\big(\frac{m_a}{5\times 10^{-12}}\big)$ for the proposed specifications of different future
    missions from  the polarized anisotropic spectral distortions of the
    CMB. In green, (the smallest contour) we plot the case with 10 times
    better sensitivity than PIXIE. The magnitude of  $g_{\gammaa}$ is
    expressed in units of $g_{10}= 10^{-10}$ GeV$^{-1}$ with the average
    signal $\bar{\mathcal{I}}= 8.3 \times 10^{-8}$ and  $1.1 \times 10^{-6}$ for $m_a= 5 \times 10^{-12}$ eV and $5 \times 10^{-13}$ eV respectively.}}\label{fisher-4}
\end{figure}
 \subsection{Comparison with other experiments}
 {We compare CMB forecasts with  with the current bounds from CAST
   experiment \cite{cast}, SN1987A \cite{Payez:2014xsa} and X-ray bounds from Coma
   cluster \cite{Conlon:2015uwa}  in Fig. \ref{fisher-5}. The $95\%$ upper
   limits are shown.  The future CMB missions can therefore provide competitive constraints
 compared with the lab experiments such as CAST.   }
Other bounds \cite{axion_2} available in the literature have  looked at the
extragalactic scenario with a much lower value of $\Ne$ and magnetic
field. {The bounds on $g_{\gammaa}B_{\rm T}$ by Tashiro et
al. \cite{Tashiro:2013yea} are obtained using the primordial magnetic field
with value today of the order nG.  Other astrophysical constraints are from
the measurement of gamma ray signal from
SN 1987A   ($g_{\gammaa}\leq 5.3
\times 10^{-12}\, \text{GeV}^{-1}$) \cite{Payez:2014xsa} and from X-ray
observations of the Coma cluster ($g_{\gammaa}\leq 1.4 \times 10^{-12}\, \text{GeV}^{-1}$) \cite{Conlon:2015uwa}.} 

 \begin{figure}
\centering
\begin{subfigure}{0.9\textwidth}
\centering

\includegraphics[width=0.9\textwidth]{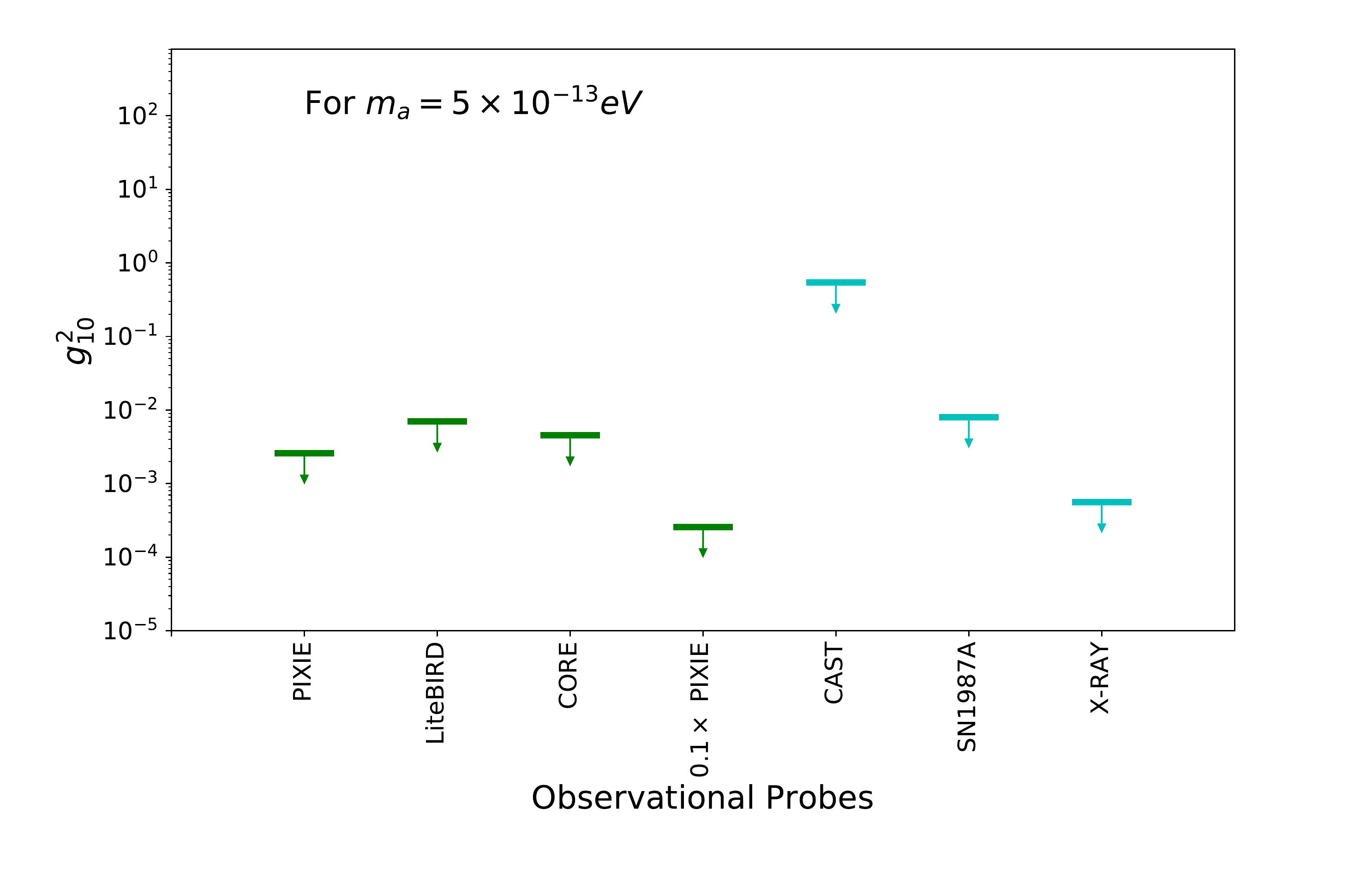}
\caption{From polarized resonance conversion}
\end{subfigure}\\
\begin{subfigure}{0.9\textwidth}
\centering
\includegraphics[width=0.9\textwidth]{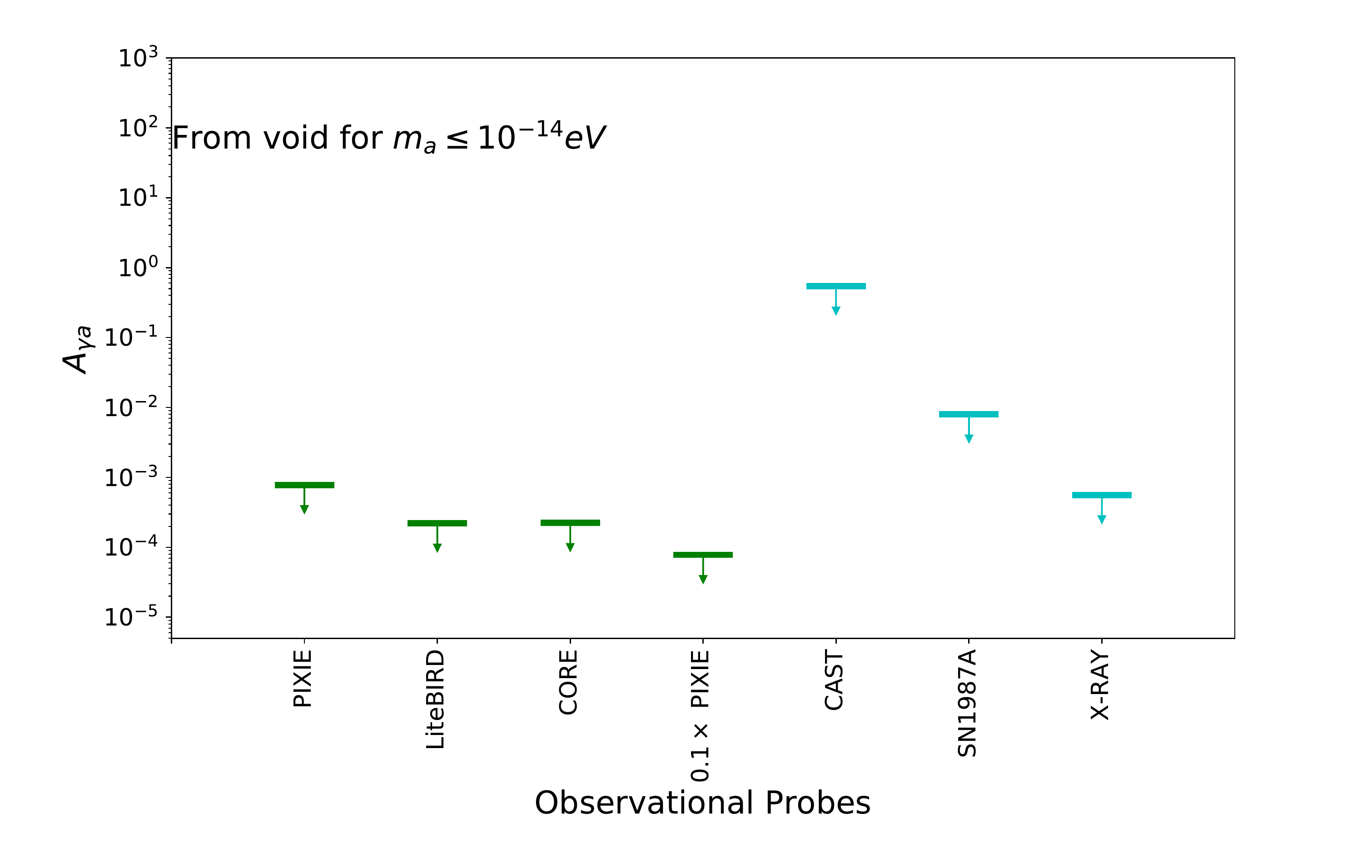}
\caption{From unpolarized non-resonance conversion}
\end{subfigure}
\captionsetup{singlelinecheck=on,justification=raggedright}
\caption{{Assuming a Gaussian probability distribution function, $2\sigma$
  upper limit achievable on (a) $g^2_{10}$ for resonant conversion in the
  Galactic magnetic field for $\ma=5\times 10^{-13}~{\rm eV}$  and (b) on
  $A_{\gammaa}\propto (g_{\gammaa}B_T^{\rm rms})^2$ for non-resonant
  conversion in voids are shown for different CMB missions  {with conservative marginalization over both dust and synchrotron.} We also plot
  the $95\%$ upper bound only on $g^2_{10}$ (cyan) from ground based
  experiment CAST \cite{cast}, the gamma ray flux of SN1987A
  \cite{Payez:2014xsa} and from X-ray observations of the Coma cluster \cite{Conlon:2015uwa}. }}\label{fisher-5}
\end{figure}

\section{Conclusions}\label{conclusion}
 We have studied a new avenue of spectral distortion of CMB photons
due to photon-ALP  {and photon-LSP} conversion in presence of our local magnetic field of
Milky way. We consider both resonant
(Sec. \ref{spectral-dist-reso}) and non-resonant (Sec. \ref{non-resonant})
Photon-ALP  {and photon-LSP} conversions.  {Even though we have done
  our calculations specifically for
  pseudoscalars such as axions, our results apply, with trivial
  correspondence between the coupling constants and rotation of the polarization
  by $90^{\circ}$, almost unchanged to the scalar
  particles.} The only observable difference between the scalar and
  pseudoscalars is that in the case where we have a polarized signal, the
  polarization of the distortion (photons which disappear due to conversion
  to scalars or pseudoscalars) is along the direction of the transverse
  magnetic field in case of pseudoscalars (Eq. \ref{Eq:int}) and in the
  orthogonal direction in the case of scalar particles. (Eq. \ref{Eq:scalar}).

  The resonant conversions can happen in the Galactic
halo for $10^{-14} ~{\rm eV} \lesssim \ma \lesssim 10^{-11}~{\rm eV}$. The probability
of conversion depends on the electron distribution as well as the large
scale magnetic field structure of the Galaxy imparting a characteristic
anisotropy to the spectral distortion. In addition this distortion is
$100\%$ polarized. The polarized anisotropic spectral distortion provides
an ideal target for future CMB missions which would focus on the polarized
signals. The anisotropic nature of the signal means that it is accessible
by the CMB experiments without absolute calibration. {The
  polarization of the signal for scalars and pseudoscalars is orthogonal to
each other. In this very interesting case, we therefore get the mass of the
particle from the anisotropy pattern, which varies with the particle mass,
and from the  polarization we can tell whether the particle coupled to photons is a scalar or a pseudoscalar
particle.}

For axion masses $\ma \lesssim 10^{-14}~{\rm eV}$ we consider non-resonant conversion
in the small scale turbulent Galactic magnetic field as well as the
primordial stochastic magnetic fields in the voids. This distortion is
unpolarized if it is the average over large number of random magnetic field
configurations. If the small scale turbulent magnetic fields in the Galaxy  are correlated
with the large scale magnetic field structure \cite{Jansson:2012rt}, then
we would expect an anisotropy similar to that shown in
Fig. \ref{Fig:prob-map}. The CMB spectral distortions from non-resonant
conversion depends sensitively on the model of turbulent magnetic fields in
the Galaxy and in the voids as well as the electron density profiles. We
have used a simplified model for Fisher matrix analysis to estimate the
level of distortion and the constraints on photon-axion coupling accessible
by these distortions. Our results are encouraging and motivate a more
detailed analysis with realistic models of voids and Galaxy in the future.

We have also shown that the strong cosmological constraints for $10^{-14} \lesssim
\ma \lesssim 5\times 10^{-13}~{\rm eV}$ claimed by \cite{Tashiro:2013yea}
are invalid. For lower ALP masses constraints were obtained on non-resonant
photon-axion conversion in stochastic magnetic fields using a toy model
with magnetic field abruptly changing direction on Mpc scales in
\cite{axion_2}. We have shown that the constraints from a more realistic
primordial magnetic field model are much weaker thus illustrating the
sensitivity
of the photon-axion conversion on the assumptions about the intergalactic magnetic fields.

We have used the mean
 signal for the Fisher matrix analysis in the case of resonant conversion
 in the Galactic halo. However as we can see from the maps in
 Fig. \ref{Fig:prob-map-reso}, there is large anisotropy in the signal with
the signal varying by many orders of magnitude over the sky. In
particular there are regions in the sky with much higher signal than the mean. We have
also not used all channels available in the CMB experiments  {(like PIXIE)} to simplify the
analysis. Using higher
frequency channels would require a more sophisticated model of dust emission
than our 2-parameter model. Using additional channels would
help improve the sensitivity and thus our constraints. Our results however
rely on the knowledge of Galactic electron distribution and magnetic
fields which we expect to improve significantly with the future radio
surveys, in particular with the Square Kilometer Array
\cite{2015aska.confE..41H} on time scales similar to the proposed CMB space
missions.

\color{black}
\textbf{Acknowledgements}
This work has been done within the Labex ILP (reference ANR-10-LABX-63)
part of the Idex SUPER, and received financial state aid managed by the
Agence Nationale de la Recherche, as part of the programme Investissements
d'avenir under the reference ANR-11-IDEX-0004-02. The work of SM and BDW are supported by the Simons Foundation. This research was supported
by SERB grant no. ECR/2015/000078 of Science and Engineering Research
board, Dept. of Science and Technology, Govt. of India. This research was
also supported by Max-Planck-Gesellschaft through the partner group
between MPI for Astrophysics, Garching and TIFR, Mumbai. This research made
use of computational resources of IAP, CCA and DTP-TIFR. RK would like to thank Max
Planck Institute for
Astrophysics, Garching for hospitality where part of this work was
done. RK would like to thank Basudeb Dasgupta and Amol Dighe for numerous
discussions and help in
understanding the flavor oscillation physics and to the former for also
reading the manuscript and making useful comments. SM would like to thank Joseph Silk for useful discussions and comments on the draft. The authors also acknowledge valuable comments from David Spergel, David Marsh, Eiichiro Komatsu and Hendrik Vogel on the paper.

\appendix
\section{Galactic magnetic field and electron density model}\label{sec:galmodel}
A model of the coherent component of the Galactic magnetic field
  in the disk and halo of Milky way was developed by Jansson et
  al. \cite{jansson}. Magnetic field in the Galactic halo at a radius $r$
  and height $l$ can be written into toroidal ($B^{\text{tor}}$) and
  poloidal ($B^{\text{pol}}$) component in terms of step function
  $L(l,h,w)= \left(1+e^{-2(|l|-h)/w}\right)^{-1}$. The toroidal
    component is separated into the north ($B_n$) and south ($B_s$) component  as
  \cite{jansson}
\begin{align}\label{mag-1}
\begin{split}
B^{\text{tor}}(r, l) = e^{-|l|/l_0} L(l, h_{disk} , w_{disk} )\times \begin{cases}&B_n\left(1-L(r,r_n,w_h)\right)\,\, l>0,\\
&B_s\left(1-L(r,r_s,w_h)\right)\,\, l<0, \end{cases}
\end{split}
\end{align}
The step function $L(l,h,w)$ goes to zero for $l\rightarrow 0, h\gg w$ and
unity at $l\gg h$. The first factor of $L$ means that we restrict the
toroidal component to outside the disk of height $h_{disk}$ and the second
factor of $L$ makes the field diminish outside a radius of $r_n,r_s$ for
the northern and southern regions of the Galaxy respectively.
\begin{align}\label{mag-1a}
\begin{split}
B^{\text{pol}}(r,l)= B_X e^{-r_p/r_X} \times \begin{cases}& \bigg(\frac{r_p}{r}\bigg), \text{with } r_p= r- |l|/\tan(\Theta_X^0) r>r_X,\\
&\bigg(\frac{r_p}{r}\bigg)^2, \text{with } r_p= \frac{rr_X^C}{r^c_X+|l|/\tan(\Theta_X^0)}\,\, r<r_X\,\, \&\\& \hspace{1.5cm} \Theta_X(r,l)= \tan^{-1}\bigg(\frac{|l|}{r-r_p}\bigg).\end{cases}
\end{split}
\end{align}
The following best fit parameters for the toroidal and poloidal components of the
magnetic field are fitted by \cite{jansson}:
$l_0= 5.3 \pm 1.6$ kpc, $r_n= 9.22 \pm 0.08$ kpc, $r_s >16.7$ kpc, $w_h=0.2
\pm 0.02$ kpc, $h_{\text{disk}}= 0.4 \pm 0.03$ kpc, $w_{disk}= 0.27 \pm
0.08$ kpc, $B_n= 1.4 \pm0.1\, \mu$G, $B_s= -1.1\pm 0.1\, \mu$G, $B_X= 4.6\,
\mu$G, $\Theta_X^0= 49 \pm 1^\circ$, $r_X^c= 4.8 \pm 0.2$ kpc, $r_X= 2.9
\pm 0.1$ kpc.

The electron density  decreases exponentially
 with increasing distance from the Galactic plane
 \cite{Cordes:2002wz,Gaensler:2008ec,Miller2013}. The electron density in the Galactic halo can be modeled as $\text{sech}^2 (|l|/H)$ \cite{Cordes:2002wz}. 
\begin{align}\label{elec-1}
\begin{split}
n_e(r,l)= n_1\bigg[\frac{\cos(\pi r/2A_{1})}{\cos(\pi R_{\bigodot}/2A_{1})}\bigg]\text{sech}^2 (|l|/H)U(r-A_1),
\end{split}
\end{align}
where, $U(x)$ is a step function. The value of vertical scale height of $H=0.95$ kpc was used by Cordes et al. \cite{Cordes:2002wz}, which was later modified to $H=1.8$ kpc by Gaensler et al. \cite{Gaensler:2008ec}  and $A_1=17\, \text{kpc}$, $n_1=0.035 \text{cm}^{-3}$. This indicates a much higher electron density at high latitudes than the previous analysis \cite{Cordes:2002wz}. Similar to the paper by Jansson et al. \cite{jansson} (which provides the model for the magnetic field), we use the model of electron density given by Cordes et al. \cite{Cordes:2002wz} with the improved model parameters from Gaensler et al. \cite{Gaensler:2008ec}.

The estimation of photon-ALP conversion depends on the model of the magnetic field and electron density in the high latitudes. In particular, for the toroidal component, the magnetic field strength drops exponentially
 with a  scale height of $l=5.3$ kpc. The electron density also decreases exponentially
 with increasing distance from the Galactic plane
 \cite{Cordes:2002wz,Gaensler:2008ec,Miller2013}. Current observations do not provide a good probe for the electron density in the Galactic halo. However, from the upcoming mission SKA \cite{2015aska.confE..41H} an accurate observation of electron density can improve the constraints on the value of $\Ne$ and $B$. In this paper, we will assume the electron density model of Cordes et al. \cite{Cordes:2002wz} with the parameter values according to Gaensler et al. \cite{Gaensler:2008ec}.

\bibliographystyle{unsrtads}
\bibliography{photon_axion_conversion}{}

\end{document}